\documentclass[twocolumn,tighten, table]{aastex631}
\usepackage{natbib}
\usepackage{amsmath}
\usepackage{listings}
\usepackage[normalem]{ulem}
\usepackage[section]{placeins}
\usepackage{color}
\usepackage{graphicx}
\usepackage{fancyvrb}
\usepackage{float}

\shorttitle{Direct formation of MBHs in fully cosmological simulations}
\shortauthors{L.~Mayer et al.}

\newcommand{\comments}[1]{} 

\newcommand{\soutPC}{\bgroup\markoverwith{\textcolor{cyan}{\rule[0.5ex]{2pt}{1pt}}}\ULon}

\newcommand\T{\rule{0pt}{2.6ex}}       
\newcommand\B{\rule[-1.2ex]{0pt}{0pt}} 

\begin{document}

\title{Direct formation of massive black holes via dynamical collapse in metal-enriched\\ merging galaxies at $z \sim 10$: fully cosmological simulations}


\correspondingauthor{Lucio Mayer}
\email{lucio.mayer@uzh.ch}

\author[0000-0002-7078-2074]{Lucio Mayer}
\affiliation{Center for Theoretical Astrophysics and Cosmology, Institute for Computational Science, University of Zurich,\\ Winterthurerstrasse 190, CH-8057 Z\"urich, Switzerland}

\author[0000-0002-1786-963X]{Pedro R. Capelo}
\affiliation{Center for Theoretical Astrophysics and Cosmology, Institute for Computational Science, University of Zurich,\\ Winterthurerstrasse 190, CH-8057 Z\"urich, Switzerland}

\author[0000-0003-4818-3400]{Lorenz Zwick}
\affiliation{Center for Theoretical Astrophysics and Cosmology, Institute for Computational Science, University of Zurich,\\ Winterthurerstrasse 190, CH-8057 Z\"urich, Switzerland}

\author[0000-0002-6462-5734]{Tiziana Di Matteo}
\affiliation{McWilliams Center for Cosmology, Department of Physics, Carnegie Mellon University, Pittsburgh, PA 15213, USA}
\affiliation{NSF AI Planning Institute for Physics of the Future, Carnegie Mellon University, Pittsburgh, PA 15213, USA}


\begin{abstract}
We present the results of the first fully cosmological hydrodynamical simulations studying the merger-driven model for massive black hole (BH) seed formation via direct collapse. Using the zoom-in technique as well as particle splitting, we achieve a final spatial resolution of 2~pc. We show that the major merger of two massive galaxies at redshift $z \sim 8$ results in the formation of a nuclear supermassive disk (SMD) of only 4 pc in radius, owing to a prodigious gas inflow sustained at 100--1000~$M_{\odot}$~yr$^{-1}$. The core of the merger remnant is metal-rich, well above solar abundance, and the SMD reaches a gaseous mass of $3 \times 10^8$~$M_{\odot}$ in less than a million years after the merger, despite a concurrent prominent nuclear starburst. Dynamical heating as gas falls into the deepest part of the potential well, and heating and stirring by supernova blastwaves,  generate a turbulent multi-phase interstellar medium, with a gas velocity dispersion exceeding 100~km~s$^{-1}$. As a result, only moderate fragmentation occurs in the inner 10--20~pc despite the temperature falls below 1000~K. The SMD is Jeans-unstable as well as bar-unstable and will collapse further adiabatically, becoming warm and ionized. We show that the SMD, following inevitable contraction,  will become general relativistic unstable and directly form a supermassive BH of mass in the range $10^6$--$10^8$~$M_{\odot}$, essentially skipping the stage of BH seed formation. These results confirm that mergers between the most massive galaxies at $z \sim 8$--10  can naturally explain the rapid emergence of bright high-redshift quasars.
\end{abstract}



\section{Introduction}\label{sec:introduction}

The origin of supermassive black holes (SMBHs) is still unclear. The existence of bright quasars at $z > 7$ points to an extremely rapid growth phase at the dawn of galaxy formation. This can be realized with either super-Eddington accretion onto relatively light BH seeds \citep[e.g.,][]{Madau_et_al_2014,Mayer_2019}, such as those originating from Pop~III stars, or with heavy BH seeds, with masses higher than $10^4$~$M_{\odot}$ forming via direct gas collapse \citep[][]{Woods_et_al_2019, Lodato_Natarajan2006}. Various works in the recent years have shown that growing large BHs in less than a billion years via super-Eddington accretion onto seeds with mass $< 10^4$~$M_{\odot}$ is difficult, because radiative feedback from active galactic nuclei (AGN) as well as clumpy gas dynamics in the nuclei of proto-galaxies prevent sustained central gas inflows \citep[][]{Lupi_et_al_2016,Sassano_et_al_2023}. Additionally, jets and outflows inhibiting further gas inflows from the larger scales \citep[][]{Regan_et_al_2019}. 

Direct-collapse scenarios postulate different pathways for both the formation of the precursor of the BH seed and the mechanism by which the precursor turns into a BH \citep[][]{inayoshi2020}. The latter involves typically the formation of a supermassive star (SMS) followed, or not, by a quasi-star, short-lived structures which rapidly collapse into SMBH seeds \citep[][]{Begelman_et_al_2006,Begelman_2010,Hosokawa_et_al_2013}. The most popular scenario for precursors relies on collapse in metal-free gas with molecular hydrogen dissociated by the Lyman--Werner background \citep[][]{Dijkstra_et_al_2006}. This requires formation to occur in halos at redshift sufficiently high ($z > 15$) that metal-free patches of the Universe still exist, and that they are in proximity of regions where proto-galaxies have already formed, with their stars producing dissociating ultraviolet (UV) photons. In these conditions, a metal-free proto-galactic disk cannot cool below 2000~K, thus avoiding fragmentation and star formation (SF), and rather generating a central gas accumulation via internal angular momentum transport due to gravitational instability \citep[][]{Lodato_Natarajan2006}. \citet{Wise_et_al_2019} have shown, using the largest cosmological hydrodynamical simulation designed to study such mechanism, that indeed a sufficiently large number of precursor proto-galaxies in atomic cooling halos with central gas inflows large enough to form SMSs at their centers is generated, so as to match the measured abundance of bright quasars at $z > 6$. \citet{Wise_et_al_2019} also found that dynamical heating, namely PdV work and shocks associated with rapid gas infall, is a major driver of central gas collapse and reduced fragmentation in atomic cooling halos, possibly as important as the dissociating radiation. They found that mergers are typically responsible for enhanced dynamical heating phases. More recently, \citet{Latif_et_al_2022} found heating by compression in cosmic web filaments to be important, suggesting that the environment of atomic cooling halos rather than dissociating radiation is crucial \citep[see also][]{Hirano2017}. In atomic cooling halos, due to their low masses ($<10^9$~$M_{\odot}$), dynamical heating cannot be as high as in more massive halos forming at later epochs. Indeed, the mass accretion rate of halos is $\dot M \sim V_{\rm vir}^{1/3} \sim M_{\rm vir}$, assuming that radial gas infall is driven by the gravitational potential well of the halo \citep[see, e.g.,][]{Mayer_Bonoli_2019}. In case of mergers, the infall rate can be close to the expectation from radial infall, since angular momentum can be removed very efficiently due to both hydrodynamical torques \citep[e.g. large-scale ram-pressure shocks;][]{Barnes2002,CapeloDotti2017,BlumenthalBarnes2018} and tidal torques \citep[due to gravitational forces; e.g.][]{BarnesHernquist1996,HopkinsQuataert2010,Capelo2015} ensuing from the non-axisymmetric mass distribution of the two interacting galaxies.

A much more extreme version of dynamical heating in atomic cooling halos occurs in the merger-driven model for direct collapse, originally presented in \citet{Mayer_et_al_2010} and subsequently developed in \citet{Mayer_et_al_2015} and \citet{Mayer_Bonoli_2019}. In such model, mergers between the most massive galaxies in place at $z \sim 10$, already enriched to solar metallicity values, produce supermassive disks (SMDs)  exceeding a billion solar masses without the need of any dissociating radiation to suppress cooling. The central SMDs, owing to their large mass, could become rapidly unstable to the radial general relativistic (GR) instability \citep[][]{Haemmerle_et_al_2019,Haemmerle_2020} and form a BH much larger than in the other direct-collapse scenarios. The possibility has also been conjectured that a fully-fledged SMS never forms, rather a proto-SMS would turn directly into a very massive BH seed of $10^7$--$10^9$~$M_{\odot}$, a formation route dubbed {\it dark collapse} \citep[][]{Mayer_Bonoli_2019,Haemmerle_et_al_2020}. This latter direct formation route, that essentially overrides the BH seed formation stage all-together, has been studied, and shown to be plausible, in the analytical model of \citet{Zwick_et_al_2023}, which adopts the central SMD found in the previously published simulations as a starting condition.

In the merger-driven scenario, the key is that collapse occurs in galaxies hosted in halos corresponding to rare 4--5$\sigma$ peaks weighing $10^{12}$~$M_{\odot}$, and which additionally undergo major mergers. The deep potential wells combined with the merger dynamics allow radial gas inflows to reach the phenomenal rate of $10^4$~$M_{\odot}$~yr$^{-1}$, as opposed to $< 1$~$M_{\odot}$~yr$^{-1}$. For that to happen, a large reservoir of cold gas has to be supplied to the galaxies, but this is known to occur naturally through cold flows impinging from the cosmic web. Cold flows at $z \sim 6$--7 have been already shown to allow a sizable BH seed to grow efficiently to the mass required to explain the high-redshift quasars if they can reach down to the galactic nucleus hosting it \citep[][]{DiMatteo_et_al_2012,Feng_et_al_2014}. In the merger-driven model, while dynamical heating produced by the inflow stifles fragmentation in the inner few parsecs, where gas is optically thick and thus cools slowly, the rapid cooling in metal-enriched gas at larger scales actually helps to achieve high inflow rates \citep[][]{Mayer_et_al_2015,Mayer_Bonoli_2019}. These conditions and outcomes  were verified in simulations of isolated gas-rich galaxy mergers, but were never scrutinized in fully cosmological simulations. Here we present the first fully cosmological smoothed particle hydrodynamics (SPH) simulations which, owing to the combination of the ``zoom-in'' and particle-splitting techniques, allow to reach resolutions of a few parsecs in a very massive galaxy forming at $z \sim 10$, and undergoing a major merger with a similarly massive galaxy near $z=8$. We will show that the conditions for the direct formation of a massive black hole via direct gas collapse are indeed met naturally in such cosmological mergers.


\section{Numerical setup}\label{sec:numerical_setup}

In this Section, we first describe the original cosmological uniform-volume and ``zoom-in'' simulations which are at the basis of our work (Section~\ref{sec:The_original_simulations}), then introduce our own set of cosmological zoom-in simulations (Section~\ref{sec:The_new_cosmological_zoomin_simulations}) and, finally, describe the new isolated runs (Section~\ref{sec:The_new_isolated_simulation}).


\subsection{The original simulations}\label{sec:The_original_simulations}

\begin{table*} \centering
\vspace{-3.5pt}
\caption[]{Particle specifications (initial conditions) for the runs MB, MBZ \citep[run 3HDCV in][]{Feng_et_al_2014}, and MBHR. Masses and lengths are given in $h^{-1}$~$M_{\odot}$ and $h^{-1}$~kpc, respectively. The number $\#_{\rm part-1D}$ is the number of particles the box is divided by, for a given level, regardless of the spatial extension of that level. This is why  $\#_{\rm part} = (\#_{\rm part-1D})^3$ only in MB (3200$^3$ = 3.2768e10), where there is no zoomed-in region (although, given the relatively small extension of the higher-resolution regions, in the lowest-resolution level we have $\#_{\rm part} \simeq 200^3$). The number $f$ is the factor by which we multiply the principal axes of the bounding triaxial ellipsoid, for a given level, to obtain the outer boundaries of that level ($\infty$ means that the level's outer boundaries are given by the cosmological box itself). Since MB is not a cosmological zoom-in simulation, there is only one level, which we denoted level 5 for an easier comparison with the other runs. The total number of DM particles is 3.2768e10, 12168003, and 274772016 for the MB, MBZ, and MBHR run, respectively. The total DM and gas mass are $1.095 \times 10^{19}$ and $2.89 \times 10^{12}$~$h^{-1}$~$M_{\odot}$, respectively, in the cosmological zoom-in runs. Note that the gravitational softening has different meanings for different runs (see Footnotes~1 and 4). In the highest-resolution region, by construction, the numbers of DM and gas particles are always the same and the baryon mass fraction is always 20.37 per cent, as in the original MB simulation.
\label{tab:particlespecs}}
\vspace{5pt}
{\small
\begin{tabular*}{0.99\textwidth}{m{28pt}m{37pt}|m{36pt}m{36pt}m{36pt}m{36pt}m{36pt}m{36pt}m{36pt}m{36pt}m{42pt}|m{50pt}}
Run &    & \multicolumn{9}{c|}{Dark matter particles (per level)} & Gas \B \\
 &          &  1 & 2 & 3 & 4 & 5 & 6 & 7 & 8 & 9 & particles \B \\
 \hline
MB  & m$_{\rm part}$ 	& - & - & - & - & 2.78e8 & - & - & - & - & 5.65e7 \T \B \\
    & $\epsilon_{\rm part}$ 	& - & - & - & - & 5 & - & - & - & - & 5 \T \B \\
	& $\#_{\rm part}$				& - & - & - & - & 3.2768e10 & - & - & - & - & 3.2768e10 \T \B \\
	& $f$					& -	 & - & - & - & $\infty$ & - & - & - & - & $\infty$ \T \B \\
	& $\#_{\rm part-1D}$			& -	 & - & - & - & 3200 & - & - & - & - & 3200 \T \B \\
\hline
MBZ	& m$_{\rm part}$ 	& 1.368e12 & 1.711e11 & 2.138e10 & 2.673e9 & 3.341e8 & 4.176e7 & 4.337e6 & - & - & 8.834e5 \T \B \\
	& $\epsilon_{\rm part}$ 	& 88.0 & 44.0 & 22.0 & 11.0 & 5.5 & 3.5 & 1.5 & - & - & 1.5 \T \B \\
	& $\#_{\rm part}$				& 7999341 & 3368 & 4172 & 37322 & 287917 & 561475 & 3274408 & - & - & 3274408 \T \B \\
	& $f$					& $\infty$	 & 5.6 & 4 & 3.6 & 3 & 2 & 1.5 & - & - & 1.5 \T \B \\
	& $\#_{\rm part-1D}$			& 200	 & 400 & 800 & 1600 & 3200 & 6400 & 12800 & - & - & 12800 \T \B \\
\hline
MBHR & m$_{\rm part}$ 	& 1.368e12 & 1.711e11 & 2.138e10 & 2.673e9 & 3.341e8 & 4.176e7 & 5.220e6 & 6.525e5 & 6.776e4 & 13803.2 \T \B \\
	& $\epsilon_{\rm part}$ 	& 47.2778 & 23.6389 & 11.8194 & 5.90972 & 2.95486 & 1.47743 & 0.738715 & 0.369358 & 0.173611 & 0.102151 \T \B \\
	& $\#_{\rm part}$				& 7997817 & 7306 & 39654 & 211626 & 262858 & 2383594 & 18429351 & 35895186 & 209544624 & 209544624 \T \B \\
	& $f$					& $\infty$	 & 8.4 & 7 & 5.6 & 4 & 3.6 & 3 & 2 & 1.5 & 1.5 \T \B \\
	& $\#_{\rm part-1D}$			& 200	 & 400 & 800 & 1600 & 3200 & 6400 & 12800 & 25600 & 51200 & 51200 \T \B \\
\hline
\end{tabular*}
\vspace{5pt}
}
\end{table*}

The original MassiveBlack (hereafter MB) run \citep[][]{DiMatteo_et_al_2012} is a uniform-volume, cosmological, SPH simulation, performed with the tree-particle-mesh-SPH code {\scshape p-gadget}, a hybrid version of {\scshape gadget2} \citep[][]{Springel_2005}, and run from redshift $z = 159$ to 4.75 (assuming the Wilkinson Microwave Anisotropy Probe 5-year -- WMAP5 -- cosmological parameters; \citealt{Dunkley_et_al_2009}: $\Omega_{\rm m} = 0.26$, $\Omega_{\Lambda} = 0.74$, $\Omega_{\rm b} = 0.044$, $\sigma_8 = 0.8$, $n_{\rm s} = 0.96$, and $h = 0.72$; see table~1 in \citealt{Feng_et_al_2014}). It covers a volume of [533.333~$h^{-1}$~cMpc]$^3$ and contains $2 \times 3200^3 = 65.536$~billion particles, half dark matter (DM) and half gas particles. The gravitational softening\footnote{Note that, in the case of the MB and MBZ (described later in this section) simulations, the physical gravitational softening, $\epsilon_{\rm phys}$, is related to the quoted gravitational softening, $\epsilon$, in the following way: $\epsilon_{\rm phys} = \epsilon/(1 + z)$ $\forall \,z$. This differs from the MBHR simulation (and its lower-resolution counterparts), as described in Section~\ref{sec:The_new_cosmological_zoomin_simulations}. Moreover, the meaning itself of gravitational softening differs amongst codes, such that a softening of 1.4~kpc in {\scshape gasoline2} is equivalent to a softening of 1~kpc in {\scshape gadget} \citep[see appendix~C in][]{Kim_et_al_2016}. The choice of softening in MB follows the rule of thumb $\epsilon \simeq L/(N \times X)$, where $L = 533.333~h^{-1}$~cMpc, $N = 3200$, and $X = 30$. This is applied to both DM and gas particles (even though they have different particle mass, since the numbers of gas and DM particles are the same: $N_{\rm gas} = N_{\rm DM}$).} of all particles is 5~$h^{-1}$~ckpc $\forall \,z$ and the particle mass for gas and DM is $5.65 \times 10^7$ and $2.78 \times 10^8$~$h^{-1}$~$M_{\odot}$, respectively.\footnote{The choice of DM particle mass follows $m_{\rm DM} = L^3 \rho_{\rm C} (\Omega_{\rm m} - \Omega_{\rm b})/N^3$, where $\rho_{\rm C} = 3 H_0^2/(8 \pi G) = 2.775 \times 10^2 h^2$~$M_{\odot}$~kpc$^{-3}$ is the critical density of the Universe, $H_0 = 100 \,h$~km~s$^{-1}$~Mpc$^{-1}$ is the Hubble constant, and $G$ is the gravitational constant. Since $N_{\rm gas} = N_{\rm DM}$, the gas-to-DM mass particle ratio is simply given by $\Omega_{\rm b}/(\Omega_{\rm m}-\Omega_{\rm b})$.}

The relatively high mass resolution of MB was sufficient to show sustained high levels of BH accretion rates, enough to explain the existence of $\sim$10$^9$~$M_{\odot}$ BHs at $z \sim 7$, caused by high gas densities produced by the steady high-density cold gas flows responsible for assembling the first galaxies in the first place. However, the spatial resolution was not sufficient to follow the cold gas inflows below sub-kpc scales.

To overcome this, \citet{Feng_et_al_2014} used the zoom-in technique, described below, to re-simulate three high-redshift halos from the original MB simulation, chosen amongst the halos with the most massive BHs at $z = 6$, all with a similar mass (a few times $10^{12}$~$h^{-1}$~$M_{\odot}$) at $z = 6$ but with different environments. In particular, their Halo~3 is in a relatively quiescent environment and without a violent history, except for a major merger at $z \sim 8$, therefore ideal for our study, since we wish to investigate the outcome of a major merger between massive high-redshift galaxies. The mass of Halo~3 is $\gtrsim 10^{12}$, $\lesssim 2 \times 10^{12}$, and $\gtrsim 2 \times 10^{12}$~$h^{-1}$~$M_{\odot}$ at $z = 7$, 6, and 4.5, respectively.

The zoom-in technique is described in \citet{Feng_et_al_2014} and we give here the details of only one of their runs, namely run 3HDCV in their table~2, which we rename here MBZ (for MB-zoom). The highest-resolution region is defined by finding the DM particles in the friends-of-friends group corresponding to the selected halo at $z = 6$ in the original MB simulation. They then find the initial (at $z_{\rm i} = 159$) positions of these DM particles and enclose them with a bounding tri-axial ellipsoid with principal axes (2$a$, 2$b$, 2$c$) = (5500, 4583.33, 5316.66)~$h^{-1}$~ckpc. The highest-resolution zoom region is defined as 1.5 times larger than such bounding ellipsoid and is populated with high-resolution gas and DM particles, of spatial resolution $\sim$3 times better than that of MB (i.e. 1.5~$h^{-1}$~ckpc) and mass resolution roughly\footnote{At each level of refinement, the spatial resolution improves by a factor of $f \sim 2$ and, consequently, the mass resolution improves by a factor of $f^3$. However, at the highest level of refinement, we include also gas particles, therefore $m_{\rm DM-highest-level} = 0.125 \times m_{\rm DM-second-to-highest-level}-m_{\rm gas}$.} 64 times better than that of MB (i.e. $m_{\rm DM} = 4.3 \times 10^6$~$h^{-1}$~$M_{\odot}$). Outside the highest-resolution region, a series of shells are populated with lower-resolution DM particles (i.e. there is no gas in the outer shells), until they reach a spatial resolution eight times worse than that of MB (i.e. the spatial resolution of the outer, ``unlimited'' shell is sixteen times worse). For more details on this and the other runs described here, see Table~\ref{tab:particlespecs}.


\subsection{The new cosmological zoom-in simulations}\label{sec:The_new_cosmological_zoomin_simulations}

We re-ran the original zoom-in simulation \citep[MBZ;][]{Feng_et_al_2014} with much higher resolution and without including any BH seeds based on sub-grid recipes, since we wish to study directly if conditions for massive BH seed formation arise in the gas flows within the merging galaxies.

We thus first re-created the initial conditions (ICs) of MBZ, using the version of the {\scshape n-GenIC} code \citep[][]{Springel_et_al_2005,Angulo_et_al_2012} described in \citet{Feng_et_al_2014}, to have a low-resolution control run (MBLR), and then constructed two additional ICs with increasing resolution (MBMR and MBHR), as described below.

The physical gravitational softening\footnote{Note that, in the case of the new simulations, the physical gravitational softening is related to the quoted gravitational softening, $\epsilon$, in the following way: $\epsilon_{\rm phys} = \epsilon$ for $0 \le z \le z_{\rm change}$ and $\epsilon_{\rm phys} = \epsilon (1 + z_{\rm change})/(1 + z)$ for $z > z_{\rm change}$, This differs from the MB and MBZ simulations. See also Footnote~1.} of the DM particles in the highest-resolution region of the simulation is set equal to (1/60) times the mean inter-particle distance. Therefore, we set $\epsilon_{\rm DM} = (1/60) (533333/N)$~$h^{-1}$~kpc (for $0 \le z \le 9$), where $N = 12800$, 25600, and 51200 for the MBLR, MBMR, and MBHR simulation, respectively (see Table~\ref{tab:particlespecs}). The gravitational softening of the gas particles is related to that of DM, such that $\epsilon_{\rm gas} = \epsilon_{\rm DM} (m_{\rm gas}/m_{\rm DM})^{1/3}$. In the MBHR case, this means that, for $0 \le z \le 9$, $\epsilon_{\rm DM} = 0.174$ and $\epsilon_{\rm gas} = 0.102$~$h^{-1}$~kpc, and the Plummer equivalent softenings for DM and gas are 0.124 and 0.073~$h^{-1}$~kpc, respectively. The minimum gas smoothing length is five per cent of the softening (i.e. 7~pc in the MBHR run). We note that, in MB and MBZ, the rule-of-thumb formula applies to the comoving softening, and the gas and DM softenings are equal to each other, hence the ICs of MBZ and MBLR are not perfectly identical.

We use the $N$-body, SPH code {\scshape gasoline2} \citep[][]{Wadsley_et_al_2004,Wadsley_et_al_2017} with a Wendland $C^2$ SPH kernel \citep[][]{Wendland_1995,Dehnen_Aly_2012,Keller_et_al_2014} with 50 neighbours (and a cubic spline kernel for the gravity); thermal energy and metal diffusion \citep[using a diffusion coefficient 0.05;][]{Wadsley_et_al_2008,Shen_et_al_2010}; the geometric-density-average force expression \citep[][]{Keller_et_al_2014}; hydrogen, helium, and metal cooling \citep[][]{Shen_et_al_2010,Shen_et_al_2013}, assuming a \citet{Haardt_Madau_2012} UV/X-ray diffuse background; SF \citep[][]{Stinson_et_al_2006}, wherein one gas particle can transform into one stellar particle (of identical mass) if the gas density and overdensity are higher than $5\, m_{\rm H}$~g~cm$^{-3}$ and 2.63, respectively, where $m_{\rm H}$ is the hydrogen mass, and the gas temperature is lower than $10^4$~K; and supernova (SN) feedback \citep[][]{Stinson_et_al_2006}. We impose two requirements on the \citet{Jeans_1902} length, defined as $c_{\rm s}[\pi/(G \rho)]^{1/2}$, where $c_{\rm s} = (\gamma P/\rho)^{1/2}$, $P$, and $\rho$ are the gas speed of sound, pressure, and density, respectively, and $\gamma = 5/3$ is the adiabatic index. The first requirement imposes that the Jeans length cannot be smaller than the gravitational softening. The second requirement is a pressure floor, $P > \alpha G [\rho \, {\rm max}(\epsilon_{\rm gas}, h_{\rm smooth})]^2$, where $\epsilon_{\rm gas}$ and $h_{\rm smooth}$ are the physical gravitational softening and smoothing length of the gas, respectively, and $\alpha$ is a safety parameter, such that the Jeans length is resolved by at least $N_{\rm l}$ smoothing lengths. We use $\alpha = 5$, so that $N_{\rm l} \simeq 5$ when $h_{\rm smooth} \ge \epsilon_{\rm gas}$ and can be as large as $\sim$100 when $h_{\rm smooth} < \epsilon_{\rm gas}$ (because $h_{\rm smooth}$ cannot be $< 0.05 \epsilon_{\rm gas}$).

\begin{figure*}
\includegraphics[width=0.50\textwidth]{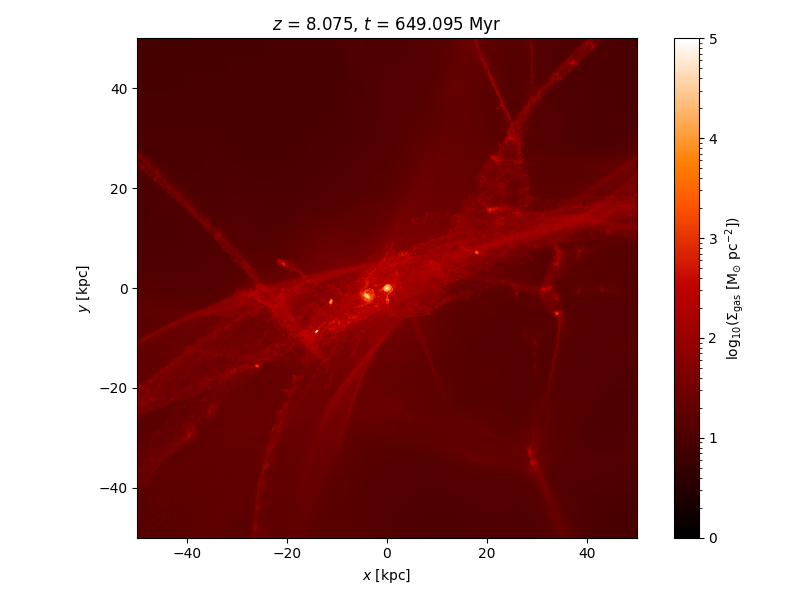}
\includegraphics[width=0.50\textwidth]{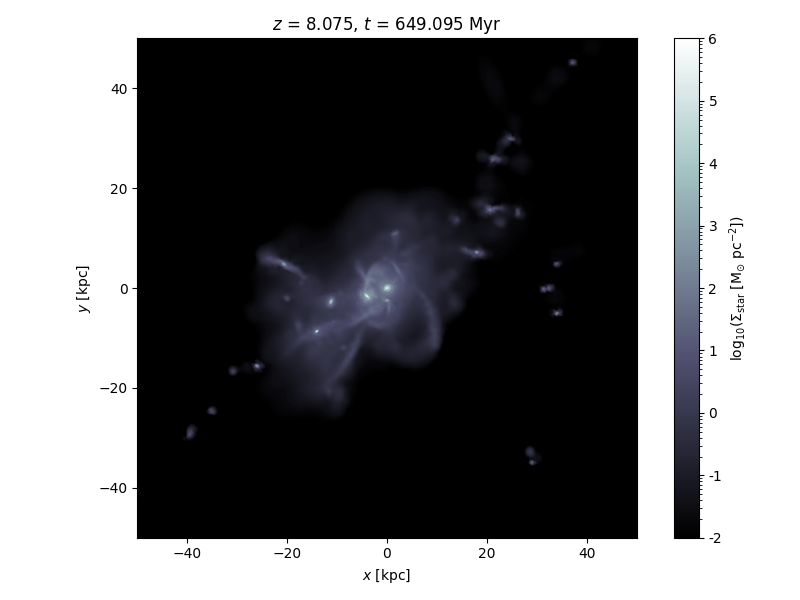}
\caption{Surface density maps of gas (left-hand panel) and stars (right-hand panel) of the central region of the high-resolution cosmological zoom-in simulation (MBHR) at $z = 8.075$, just before we cut out the region outside the virial radius of the primary system ($r_{\rm vir} = 38.996$~kpc), perform particle splitting, and start the isolated simulation.}
\label{fig:MBHR_maps_100kpc}
\end{figure*}

The initial (i.e. at $z_{\rm i} = 159$) temperature of all gas particles is equal to $T_{\rm i} = T_{\rm CMB} (1 + z_{\rm i})^2 /(1 + z_{\rm t}) = 429.9$~K, where $T_{\rm CMB} = 2.726$~K is the cosmic microwave background (CMB) temperature and $1 + z_{\rm t} \simeq 160 (\Omega_{\rm b}h^2/0.022)^{2/5}$ \citep[][]{Loeb_2008}. The initial density of gas particles is instead approximately given by $(\Omega_{\rm b}/\Omega_{\rm m}) \rho_{\rm crit}(z_{\rm i})$, where $\rho_{\rm crit}(z_{\rm i}) = 2.775 \times 10^4 h^2 [\Omega_{\rm m}(1 + z_{\rm i})^3 + \Omega_{\Lambda}]/(8 \pi G) \simeq 2.775 \times 10^4 h^2 \Omega_{\rm m}(1 + z_{\rm i})^3/(8 \pi G) \simeq 1.0377 \times 10^{-23}$~g~cm$^{-3}$.

We use the {\scshape amiga halo finder} code \citep[version 1.0-0.101;][]{Gill_et_al_2004,Knollmann_Knebe_2009} to select the halos. We checked for contamination from low-resolution DM particles (i.e. all DM particles initially outside the highest-resolution region) at $z = 6.5$. We found that, when centered on the center of the main halo, the spherical region with zero low-resolution DM particles has a radius equal to seven (MBLR and MBMR) or ten (MBHR) times the virial radius of the main halo.

\subsection{Ultra high-resolution simulations using particle splitting} \label{sec:The_new_isolated_simulation}

The major merger in the high-resolution cosmological zoom-in simulation MBHR starts at $z \sim 9.8$ (when the virial spheres start to overlap and the ratio between the virial masses is 1 : 1.2) and ends at $z \sim 7.5$ (when the two central galaxies are not distinguishable any longer). At $z = 8.075$ ($t = 649.1$~Myr after the Big Bang; see Figure~\ref{fig:MBHR_maps_100kpc}), we carve out a subset of the high-resolution volume containing the two merging galaxies and continue the computation, which we refer to as ``isolated simulations'' hereafter. At this time, the two central structures are separated by 7.47~kpc. We take a sphere centered on the center of the main halo, of radius equal to its virial radius, and delete all particles external to it.\footnote{Leaving 11103248 DM, 7216457 gas, and 3921243 star particles.} At $r_{\rm vir} = 38.996$~kpc, the dynamical time, defined as $t_{\rm dyn} = (G \rho)^{-1/2} = [4 \pi r^3/(3 G M)]^{1/2}$ (where $M$ is the enclosed mass within $r$), is 208.6~Myr, much longer than the time between $z = 8.075$ and $\sim$7.5 ($\sim$60~Myr), when the merger ends in the MBHR simulation.\footnote{If we define $t_{\rm dyn} = 2 \pi r / v_{\rm circ}$, with $r_{\rm circ} = (GM/r)^{1/2}$, we obtain an even longer dynamical time of 648.7~Myr at $r_{\rm vir}$.} We then re-sample the (parent) gas particles by splitting them into eight (children) particles \citep[][]{Roskar_et_al_2015}, randomly distributed within the SPH smoothing kernel around each parent particle, of mass equal to 1/8 of the mass of the parent particle and the same velocity of the parent particle. The new gravitational softening of the gas particles is 7~pc (i.e. $\sim$1/20 of the MBHR softening).\footnote{The new numbers for gas particles are: total number 57731656, softening 7~pc, and mass $\sim$$1.7 \times 10$$^3$~$h^{-1}$$M_{\odot}$.} The only exception is the temperature: the children particles receive the same temperature of the parent particle. This is because the scatter kernel interpolation of the temperature leads to an overall increase in gas temperature. We then ran this first ``isolated simulation'' (MB with particle splitting and $\epsilon_{\rm gas} = 7$~pc: MBPS7pc), wherein we assumed a fixed background equal to the \citet{Haardt_Madau_2012} background at $z = 7.5$, when the merger is supposed to end. We have also changed the temperature ($10^3$~K) and density ($10^3$~a.m.u.~cm$^{-3}$) thresholds for SF. At $t = 701.9$~Myr, we further reduced the gas gravitational softening (from 7 to 2~pc) and continued the so-called MBPS2pc simulation for another few Myr.

\begin{figure*}
\includegraphics[width=0.33\textwidth]{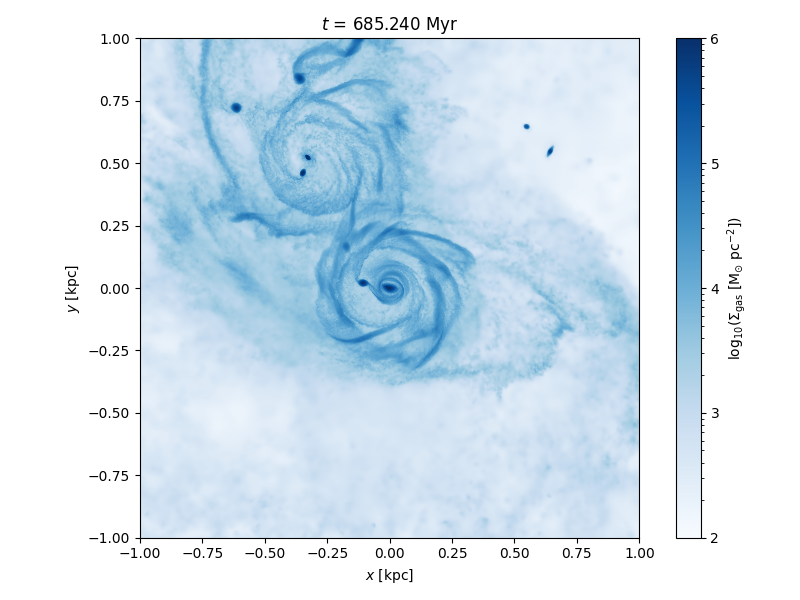}
\includegraphics[width=0.33\textwidth]{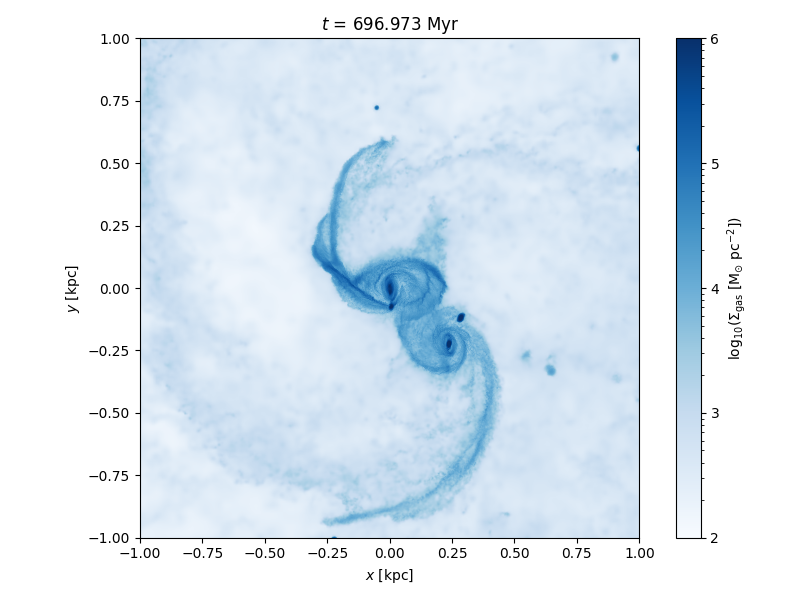}
\includegraphics[width=0.33\textwidth]{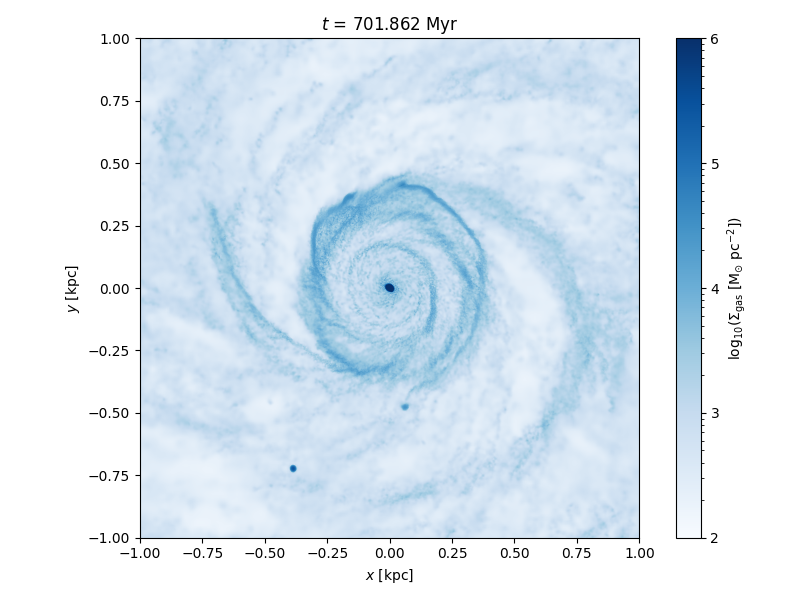}
\caption{Face-on gas density maps showing the evolution of the two galactic cores in the MBPS7pc simulation, starting when they are approaching each other for the first time with a separation of less than a kpc until the actual merger. At the initial time, the two galaxies have already merged except for their two nuclei.}
\label{fig:MBPS_maps_2kpc}
\end{figure*}


\section{Results}\label{sec:results}

\subsection{General properties of the merging galaxies in the cosmological zoom-in simulation}

The selected galaxies are highly star-forming systems. Before the final stage of the merger begins, at $z \sim 8$, the stellar mass of the most massive of the two galaxies that will merge to form the system that we selected is $\sim$$7 \times 10^{10}$~$M_{\odot}$ within $r_{\rm vir}$, $\sim$$5 \times 10^{10}$~$M_{\odot}$ within $0.1\,r_{\rm vir}$, and $\sim$$4 \times 10^{10}$~$M_{\odot}$ within the galactic disk, which is defined as a central cylinder of radius 5~kpc and half-height 0.5~kpc. We remark that these masses are already comparable to those of massive spirals at $z = 0$. Yet these galaxies are much more compact than their low-redshift counterparts, having a size of only 1.5--2~kpc if we define their size as the distance out to which a rotationally supported disk can be identified. In both gas and stellar maps of the main system at $z \sim 8$, the central disk has a sharp edge at a radius of $\sim$0.7~kpc, in agreement with other recent extremely high-resolution simulations of galaxies at $z > 5$ \citep[e.g.,][]{Fiacconi_et_al_2017,Tamfal_et_al_2022}. The galaxies host massive gas-rich disks with a  gas fraction -- defined as $M_{\rm gas}/(M_{\rm gas}+M_{\rm star})$ -- in the primary's central disk at $z \sim 8$ of $\sim$0.5. The gas disks are marginally \citet{Toomre_1964} unstable; the primary system has $Q > 1$ for $R \gtrsim 0.3$~kpc, and is $< 1$ otherwise. Indeed, the primary disk develops a strong spiral structure and a bar-like distortion, which begins to funnel gas in the nuclear region even before the merger occurs (Figure~\ref{fig:MBPS_maps_2kpc}; middle panel). Our main system's specific SF rate (sSFR) is $\sim$3~Gyr$^{-1}$ at $z = 6.5$, somewhat higher than the $\lesssim 1$~Gyr$^{-1}$ value for $z \sim 6$--7 systems observed by \citet{Salmon_et_al_2015}, but factors of a few smaller than that of extreme starburst galaxies detected at $z > 6$ \citep{Riechers_et_al_2013}. At the same redshift, we measure a stellar mass of $\sim$$2 \times 10^{11}$~$M_{\odot}$. Such properties are also consistent with the average properties of similarly massive galaxies in similarly massive halos in the BlueTides simulation \citep[][]{Wilkins_et_al_2017}, in spite of the fact that our simulation does not include AGN feedback, a potential regulator of SF. In summary, the properties of the galaxies involved in the merger are by no means exceptional, and the same applies for the galaxy merger remnant (not discussed in this paper).

\begin{figure*}
\includegraphics[width=0.24\textwidth]{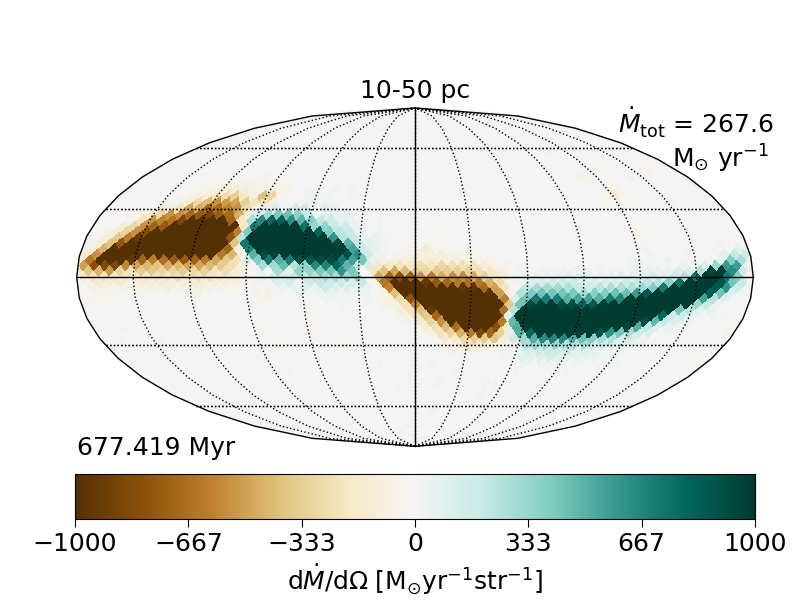}
\includegraphics[width=0.24\textwidth]{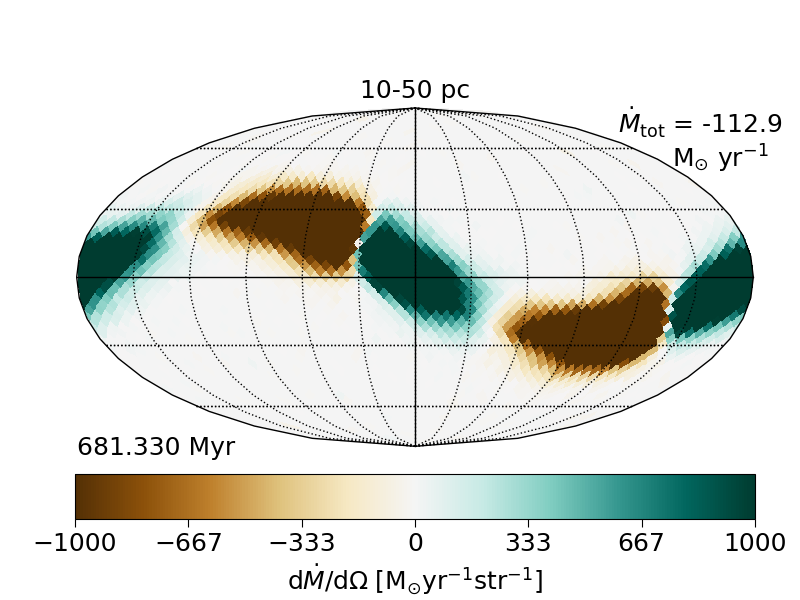}
\includegraphics[width=0.24\textwidth]{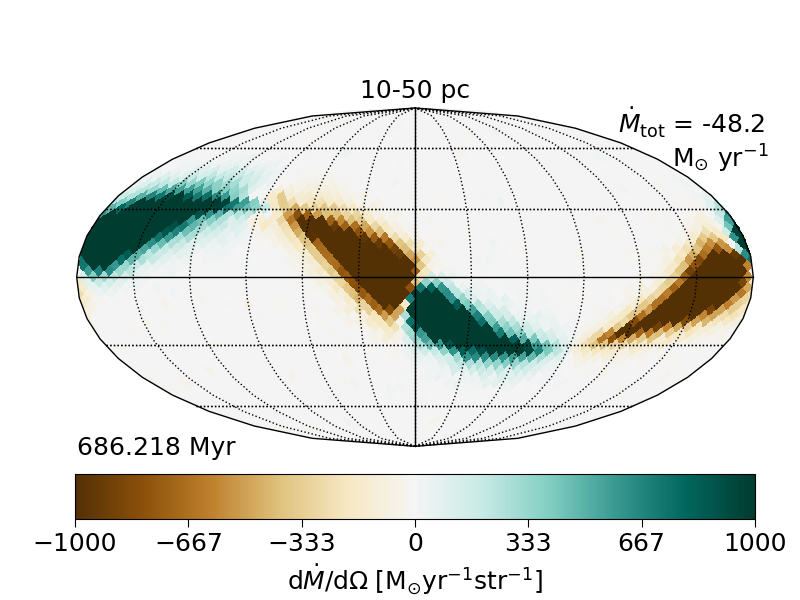}
\includegraphics[width=0.24\textwidth]{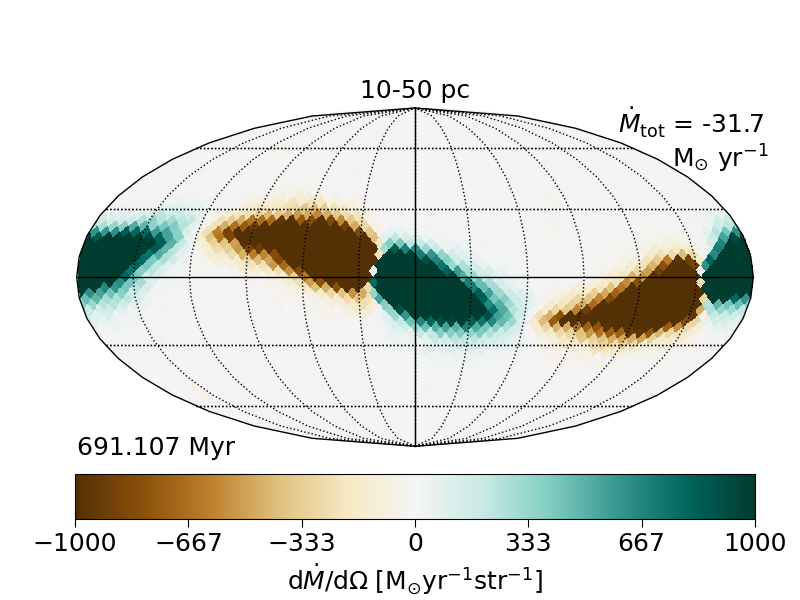}
\includegraphics[width=0.24\textwidth]{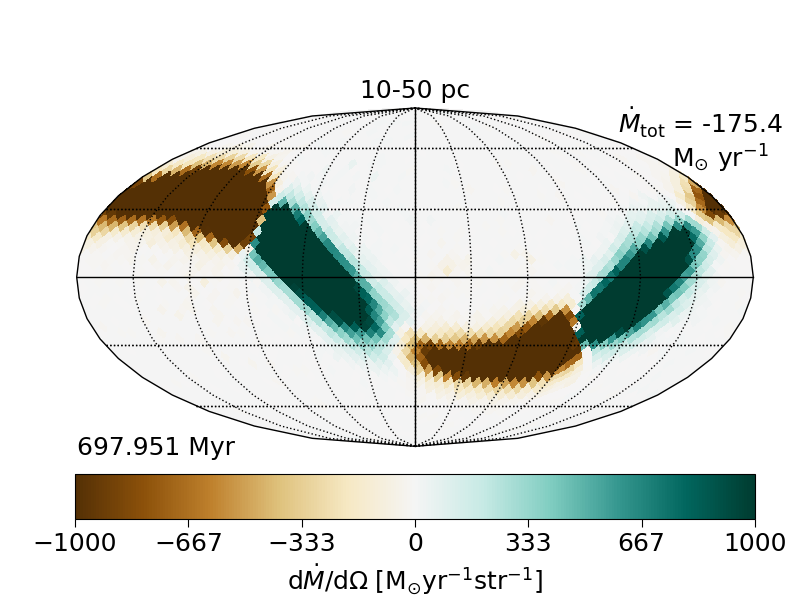}
\includegraphics[width=0.24\textwidth]{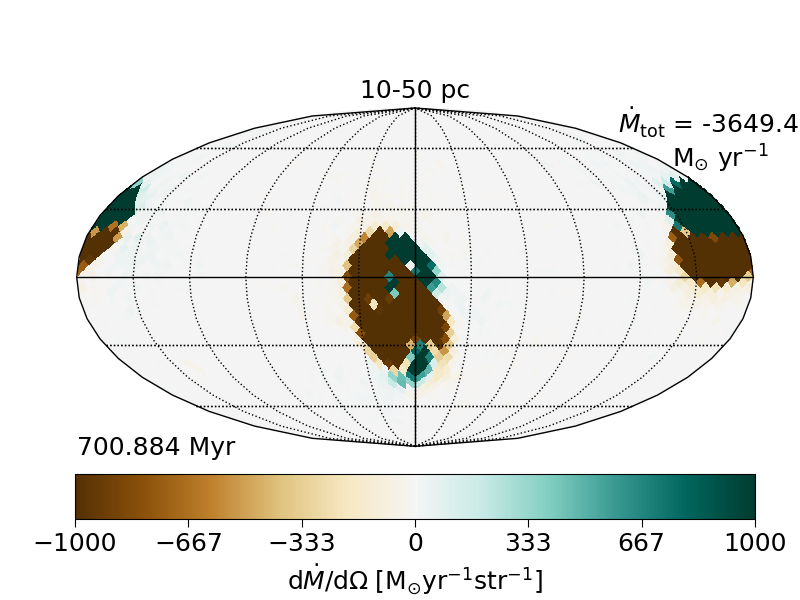}
\includegraphics[width=0.24\textwidth]{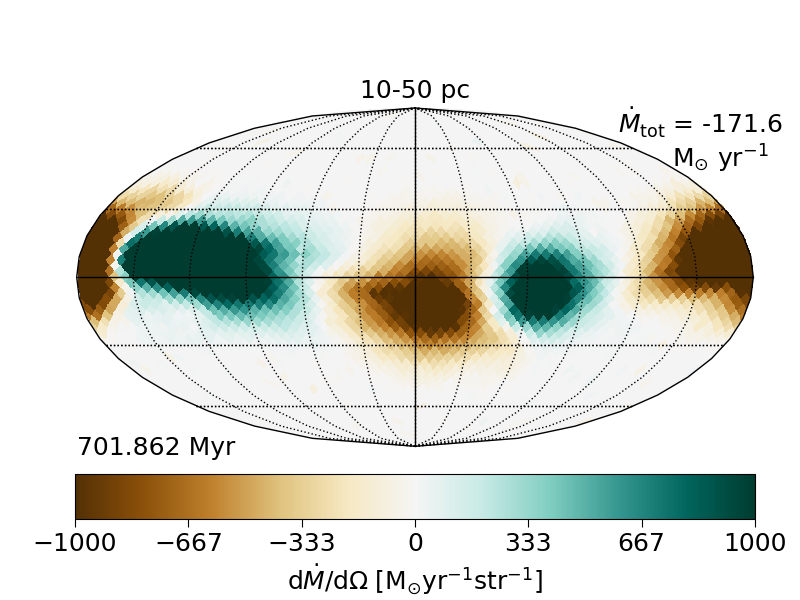}
\caption{Mollweide inflow/outflow maps for the MBPS7pc run for 10--50~pc shells at different times. The angular mass flux is shown, color-coded in magnitude according to its sign, calculated relative to the center of mass of the primary system. The total net mass flux is also shown, which is obtained summing up the different inflow and outflow
regions. The two cores are still distinguishable at $t = 700.9$~Myr (second-to-last panel) and have merged by $t = 701.9$~Myr (last panel).
}
\label{fig:mollweide_10-50pc}
\end{figure*}

\begin{figure*}
\includegraphics[width=0.24\textwidth]{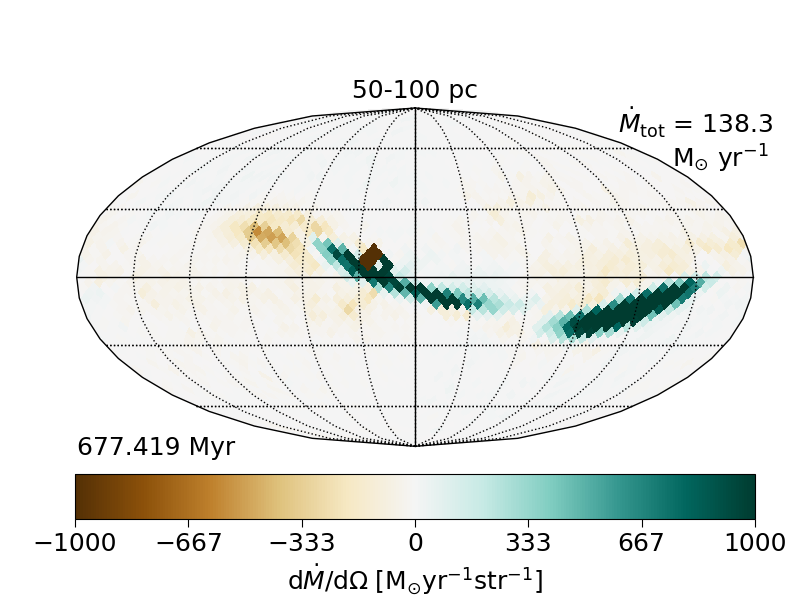}
\includegraphics[width=0.24\textwidth]{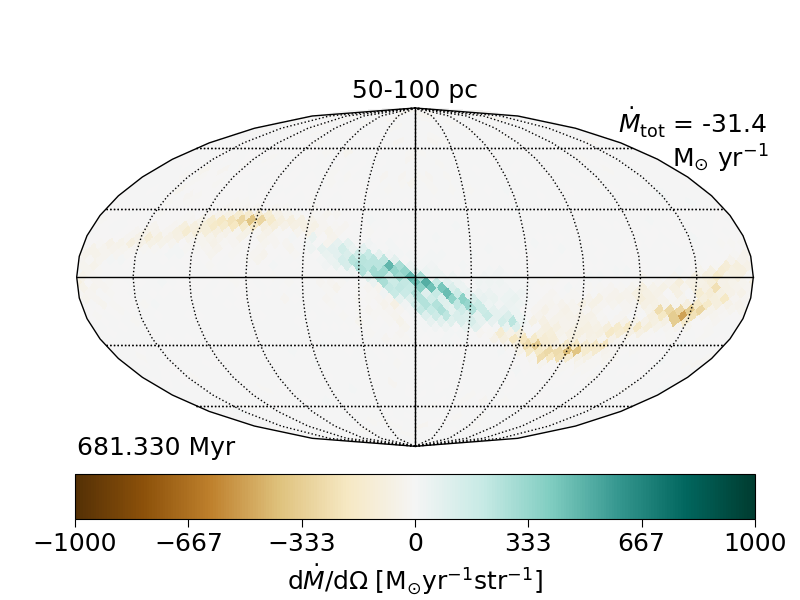}
\includegraphics[width=0.24\textwidth]{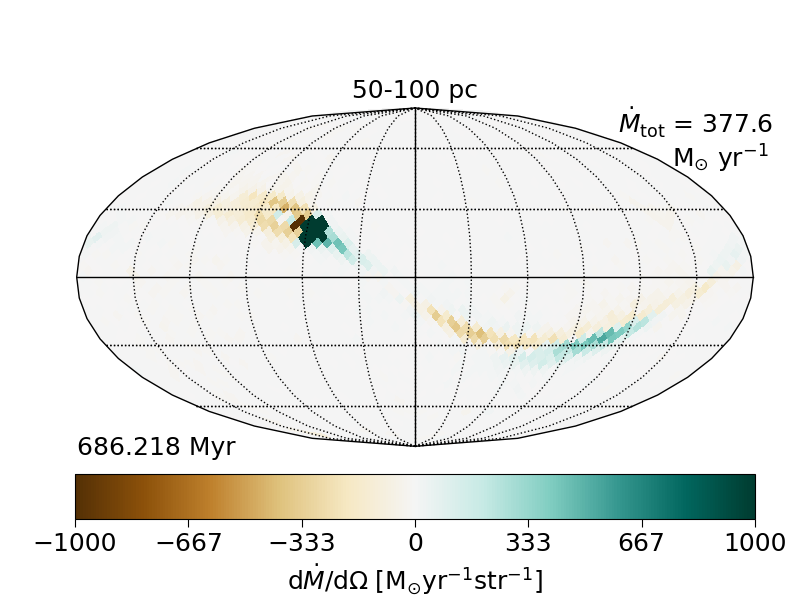}
\includegraphics[width=0.24\textwidth]{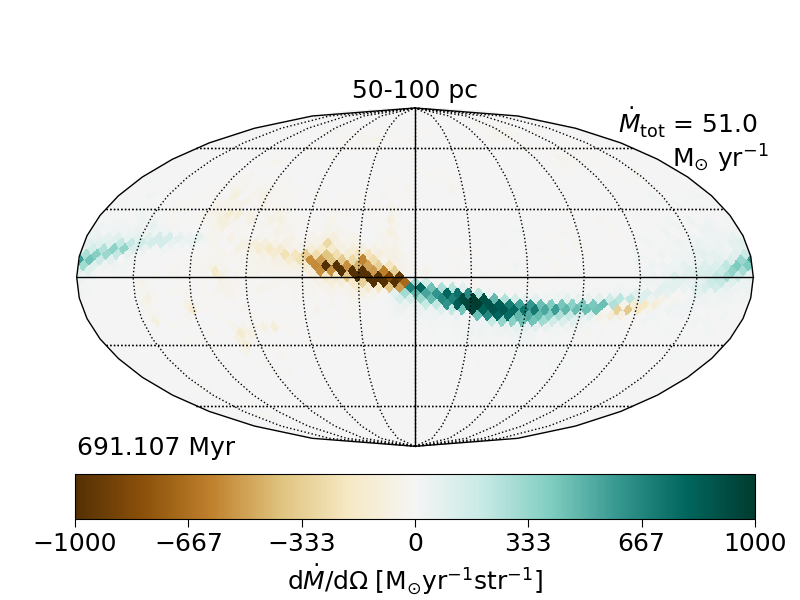}
\includegraphics[width=0.24\textwidth]{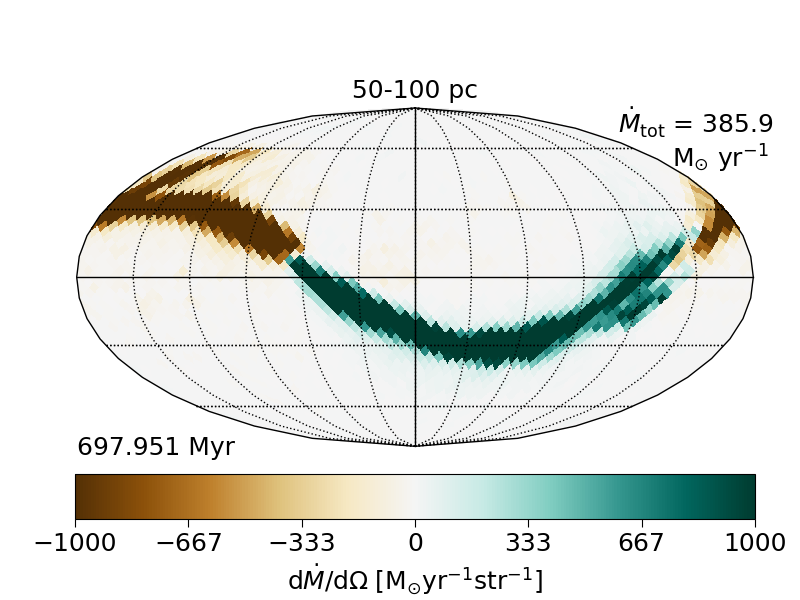}
\includegraphics[width=0.24\textwidth]{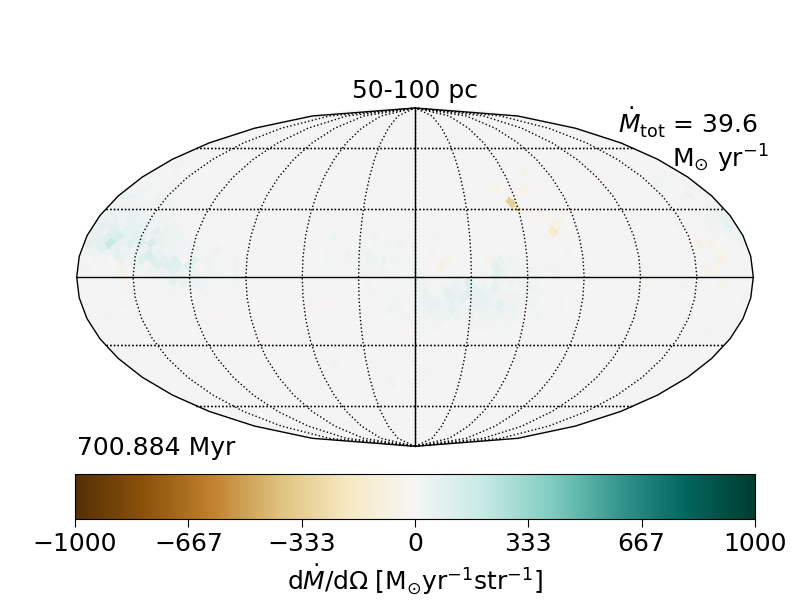}
\includegraphics[width=0.24\textwidth]{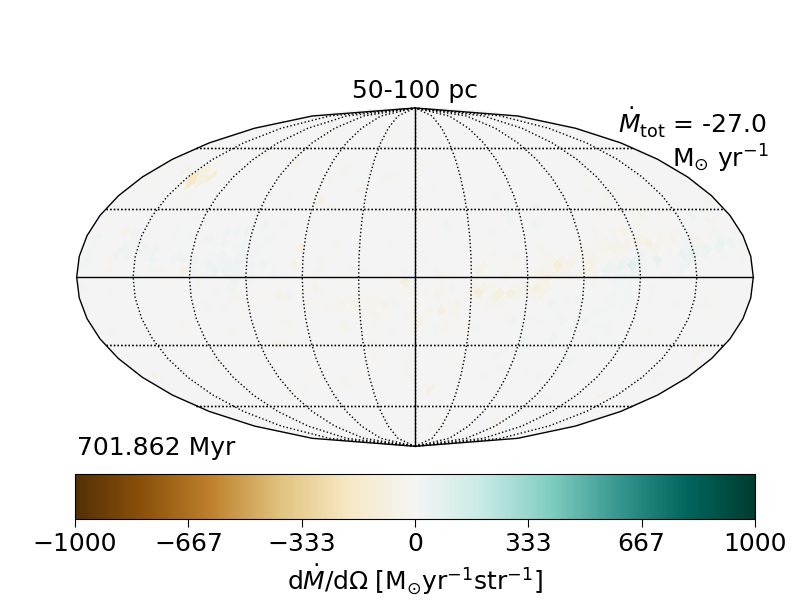}
\caption{Same as Figure~\ref{fig:mollweide_10-50pc}, but for 50--100~pc shells.
}
\label{fig:mollweide_50-100pc}
\end{figure*}

\begin{figure}
\centering
\includegraphics[width=0.49\textwidth]{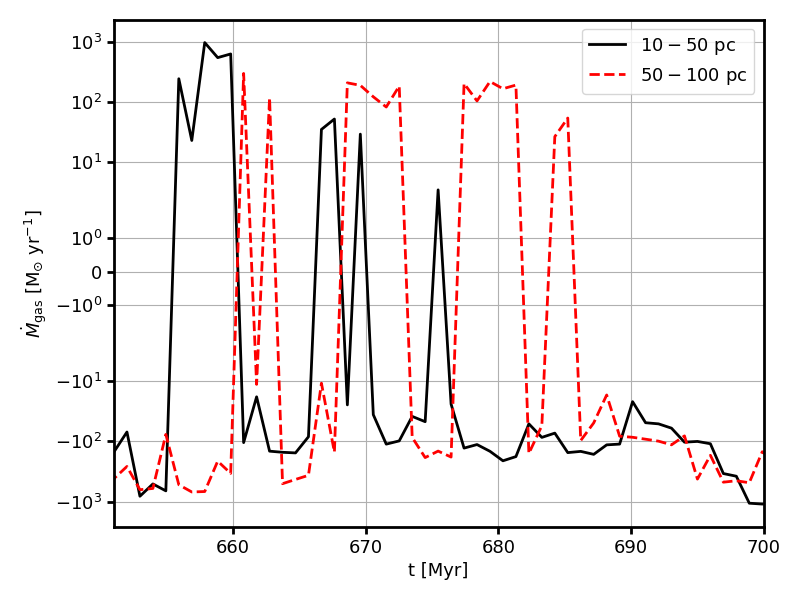}
\caption{Temporal evolution of the gas accretion rate (smoothed in time) for 10--50 pc and 50--100~pc shells centered on the primary system, from 650 to 702~Myr, in the MBPS7pc simulation. Note that, after a fluctuating behaviour, a clear net negative mass flux, which corresponds to a net gas inflow between a few hundred and a thousand solar mass per year, occurs towards the end, close to the time of the merger between the two cores.}
\label{fig:gas_accretion_smooth}
\end{figure}

\subsection{Multi-scale gas inflows and supermassive disk formation in the ultra-high-resolution particle splitting simulations}

The re-simulation with particle splitting (and $\epsilon_{\rm gas} = 7$~pc; MBPS7pc) begins when the two galactic nuclei are still {7.47} kpc apart. The structure of the galactic cores resembles rotationally supported circumnuclear disks (CNDs), confirming the results of pioneering multi-scale simulations of proto-galaxies \citep[][]{Levine_et_al_2008}, and in agreement with pc-scale binary merger simulations of high-redshift massive galaxies \citep[][]{Mayer_et_al_2015}.

The CNDs residing in the galactic cores still complete a few orbits around each other before the final coalescence, which takes about 60~Myr from the start of the re-simulation (see Figure~\ref{fig:MBPS_maps_2kpc}). As they cross pericenter passage a few times, the two galactic cores are tidally compressed, as well as stripped along the way. Tidal compression is accompanied by tidal torques that sustain radial gas inflows. There are more than $10^9$~$M_{\odot}$ contained within the two CNDs before the final merger within 20--30~pc.

The gas inflow rate is strongly time-dependent and anisotropic at all scales, as shown by the Mollweide projections of Figures~\ref{fig:mollweide_10-50pc}--\ref{fig:mollweide_50-100pc}, which show the angular mass flux in two different shells (10--50 and 50--100~pc), obtained by computing for each gas particle in a given shell and a given steradian its mass and radial velocity. During pericenter passages between the two galactic cores, it hovers around hundred solar masses per year at scales of  tens of parsecs, but can exceed 3000 solar masses per year when the two cores merge  (Figure~\ref{fig:gas_accretion_smooth}). Note that, at smaller scales, the inflow continues for slightly longer because, being due to the global torques on the gas, it is first triggered at larger scales (compare the last panels of Figures~\ref{fig:mollweide_10-50pc} and \ref{fig:mollweide_50-100pc}). The  mean gas inflow rates in the inner $\sim$10--100~pc are shown in Figure~\ref{fig:gas_accretion_smooth}. There are significant oscillations between inward and outward episodes of mass transport, reflecting the complex action of gravitational torques acting on multiple scales (e.g., spiral density waves transport mass both inwards and outwards as they can generate both positive and negative torques on parcels of fluid located at different distances). However, a net negative inflow dominates towards the end of the merger, reaching more than a few thousand solar masses per year. As the two nuclei finally collide, the largest radial gas inflow is generated at about 45~Myr after the beginning of the isolated simulation, as shown by the steep increase in the net inward radial flow in the mean accretion rate (see Figure~\ref{fig:gas_accretion_smooth}). This final episode results in the emergence of a dense compact SMD, as seen in the right-hand panel of Figure~\ref{fig:MBPS_maps_2kpc}. The central inflow rate at this point has decreased dramatically, to only a few solar masses per year, as the system has achieved a more axisymmetric configuration once again.

\begin{figure*}
\includegraphics[width=0.33\textwidth]{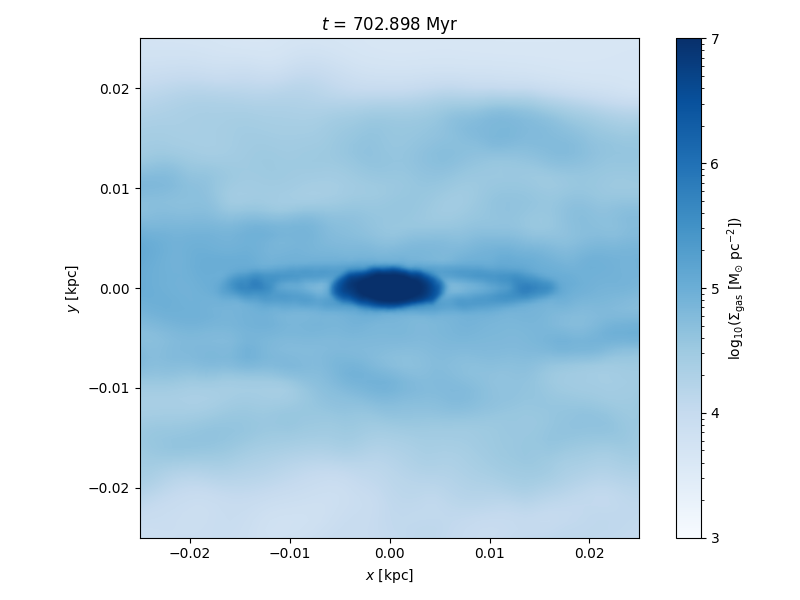}
\includegraphics[width=0.33\textwidth]{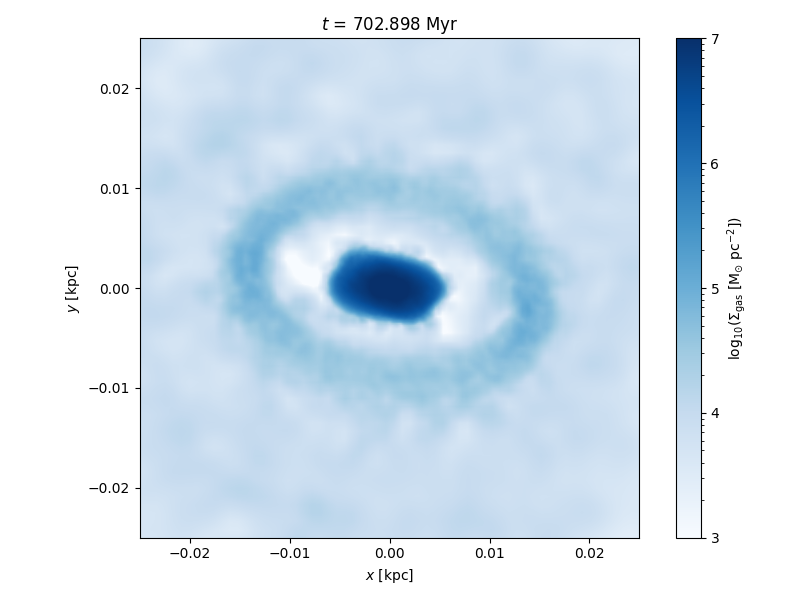}
\includegraphics[width=0.33\textwidth]{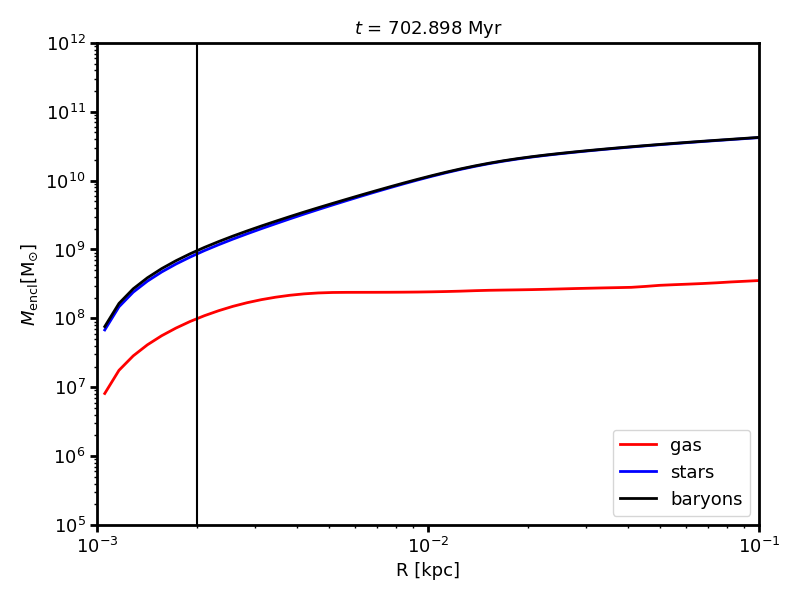}
\includegraphics[width=0.33\textwidth]{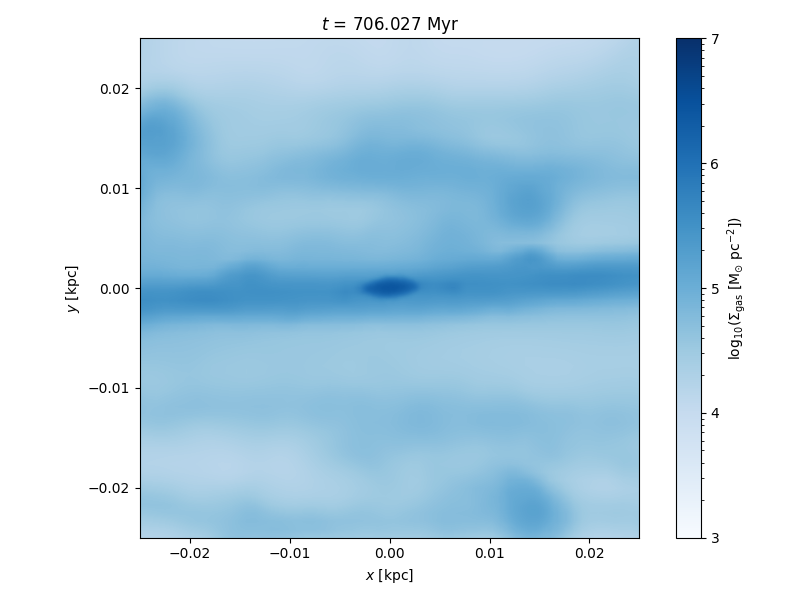}
\includegraphics[width=0.33\textwidth]{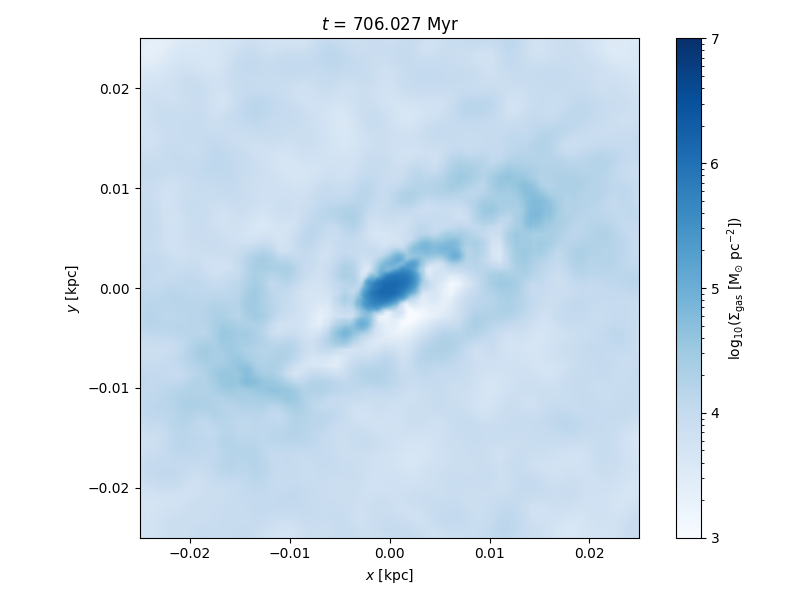}
\includegraphics[width=0.33\textwidth]{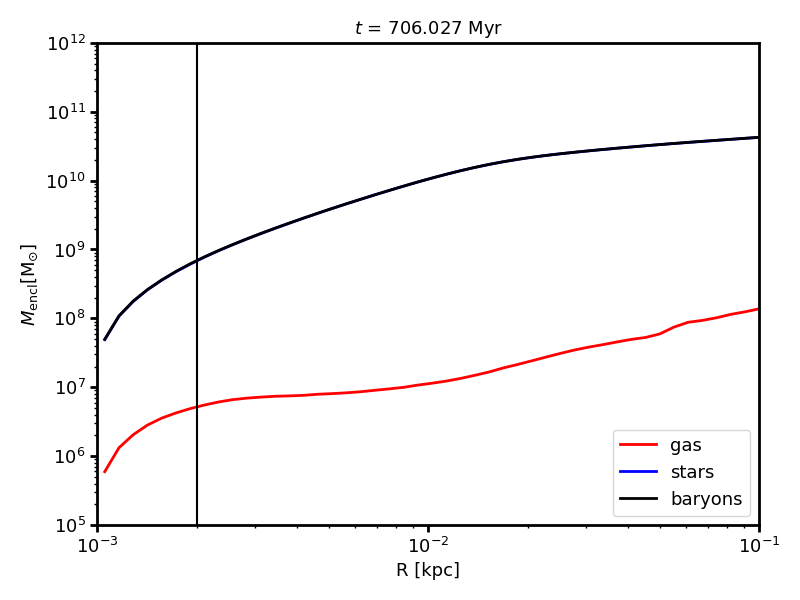}
\caption{Top panels: MBPS2pc edge-on (left-hand panel) and face-on (central panel) gas surface density maps, and enclosed baryonic mass (right-hand panel) at $t = 702.9$~Myr, soon after the merger (which occurred at 701.9~Myr, at the end of the MBPS7pc simulation). Bottom panels: same as above, but for a later time, at 706.0~Myr, after SF has consumed most of the gas. The vertical lines mark the softening length of 2~pc.} 
\label{fig:MBPS2p_gas_map_and_Mencl}
\end{figure*}

Over the course of the evolution of the two galactic cores, the high-density gas fragments, being Toomre unstable, but the inner cores of the two galaxies, within their central 10--20~pc, remain smooth, with only sporadic fragmentation, until the end. The central SMD that forms at the end in the merger remnant is about 12~pc in radius, as opposed to about 2~pc in the \citet{Mayer_et_al_2015} and \citet{Mayer_Bonoli_2019} simulations. In the latter, however, the gravitational softening was a factor of 70 smaller, being of order 0.1~pc, which would be excessively computationally expensive for the more complex multi-scale simulations presented here. Nevertheless, we restarted the simulation at $t = 701.9$~Myr, right after the merger of the two cores has been completed, using a smaller softening of 2~pc (the MBPS2pc simulation). This resulted in a significantly  smaller disk, of roughly 4~pc in size (Figure~\ref{fig:MBPS2p_gas_map_and_Mencl}), which, however, has still shrunk proportionally to the softening. This suggests that the final gas inflow and resulting SMD compactness are still artificially suppressed by the softening of the potential. Nevertheless, in the latter simulation, the SMD reaches a peak gas mass of $3 \times 10^8$~$M_{\odot}$ within a radius of about 4~pc (Figure~\ref{fig:MBPS2p_gas_map_and_Mencl}), which is within an order of magnitude of the enclosed mass obtained by \citet{Mayer_et_al_2015} in the binary merger simulations with 0.1~pc gravitational softening. We will show below that such enclosed mass is sufficient to generate an unstable SMS that could collapse directly into an SMBH. Note that the baryonic mass in the same region is higher, exceeding $10^9$~$M_{\odot}$. The large stellar mass is the product of an overly powerful starburst which lacks self-regulation due to radiative stellar feedback (mainly ionizing radiation from massive stars, not SNae which explode on timescales longer than the few Myr needed for the SMD to form); it is thus conceivable that, with a more realistic sub-grid model of stellar feedback, the enclosed gas mass would be closer to that of the binary merger simulations of \citet{Mayer_et_al_2015}. We discuss the importance of the missing radiative heating in the last section.

\begin{figure*}
\includegraphics[width=0.5\textwidth]{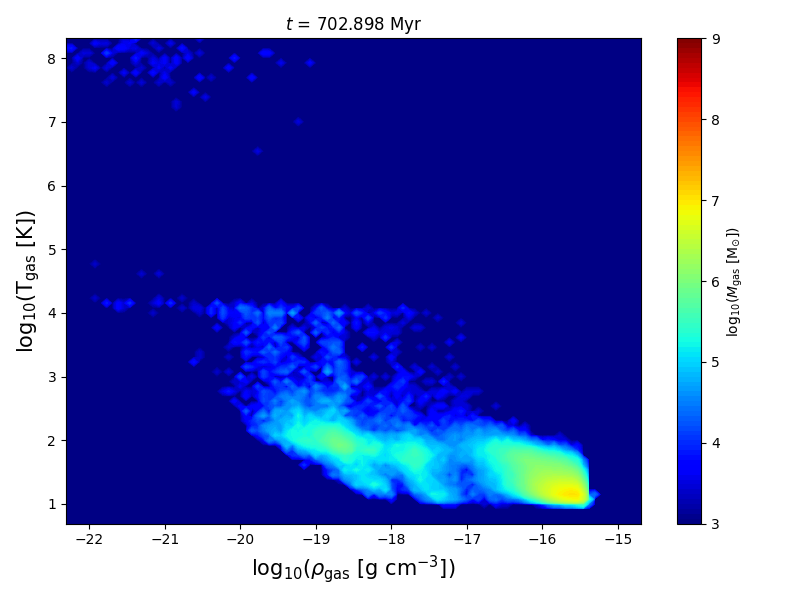}
\includegraphics[width=0.5\textwidth]{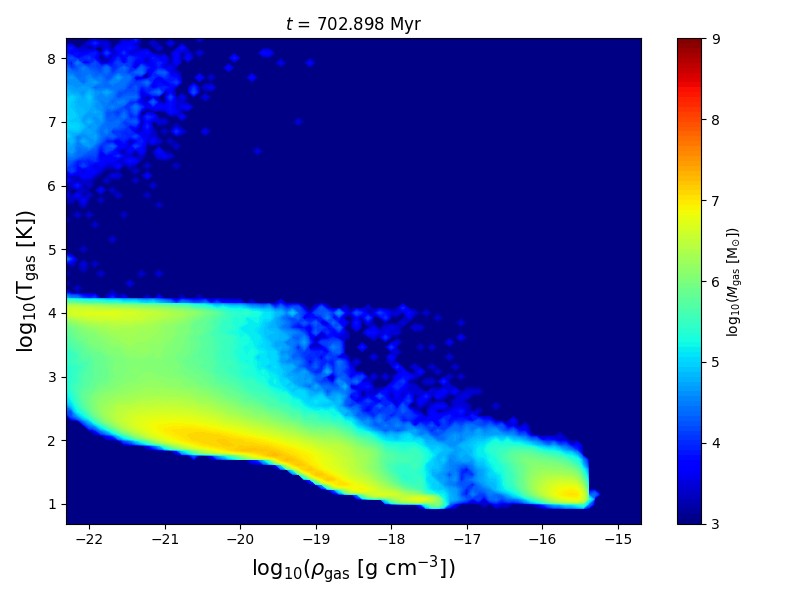}
\includegraphics[width=0.5\textwidth]{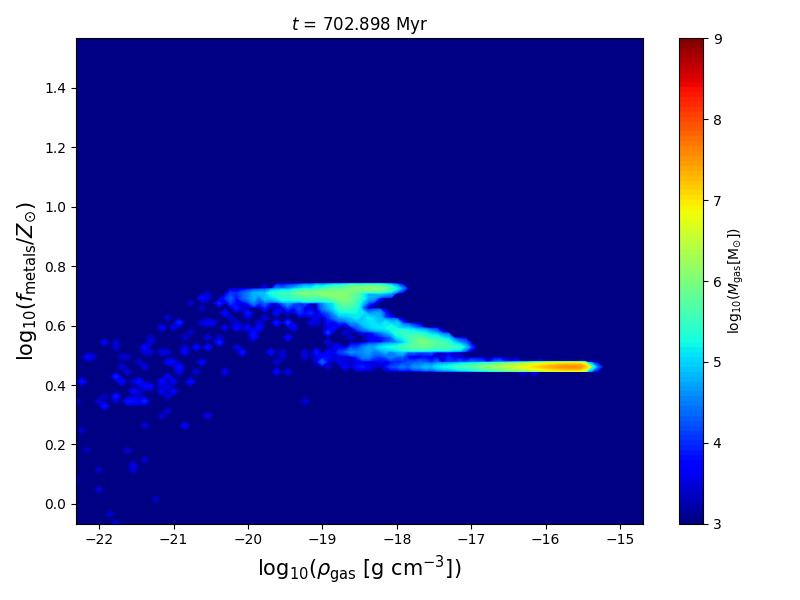}
\includegraphics[width=0.5\textwidth]{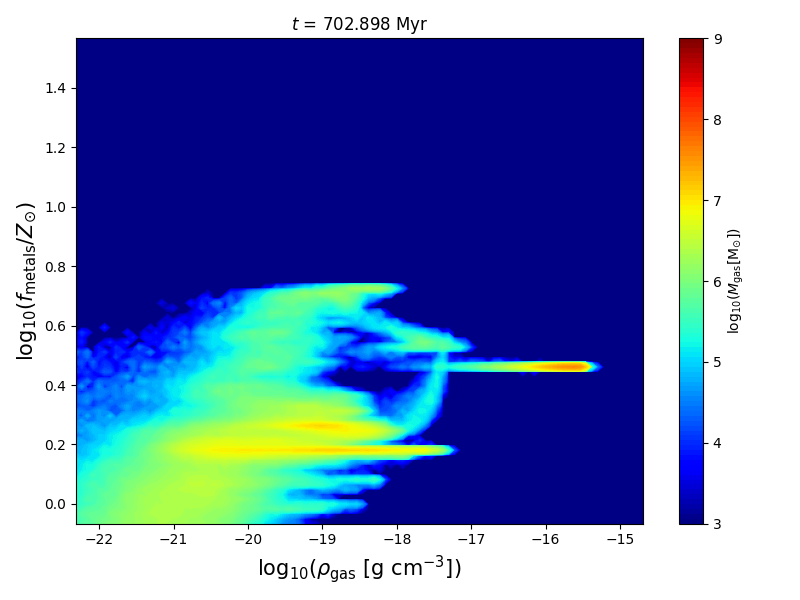}
\caption{Top panels: temperature-density diagram of gas particles within 50~pc (left-hand panel) and within the central kpc (right-hand panel), soon after the merger (in the MBPS2pc simulation), namely at the same time used to show the SMD structure in the top panels of Figure~\ref{fig:MBPS2p_gas_map_and_Mencl}. Bottom panels: same as the top panels, but for the metallicity-density diagram. It is evident that the highest-metallicity gas is in the SMD (the reddish finger-like feature at high densities), approaching ten times the solar value. The metallicity is near or above solar in the entire central kiloparsec region.}
\label{fig:phase_diagrams}
\end{figure*}

A notable difference with the binary merger simulations is that SF is more vigorous even in the nuclear region of the galaxies, both before and during the final merger phase, reaching rates, within 10--50~pc, as high as 400~$M_{\odot}$~yr$^{-1}$ at the last pericentric passage. SN feedback generates bubbles and holes in the interstellar medium (ISM), leading to a much more disturbed, patchy, and multi-phase ISM. In Figure~\ref{fig:phase_diagrams}, we show the phase diagrams in the isolated simulation with $\epsilon_{\rm gas} = 2$~pc. The warm/hot medium generated by PdV work and feedback heating is present in both, but a much colder and denser phase, with $T < 1000$ K, is present only at the much higher resolution of the particle-splitting simulation. The SMD is comprised primarily of such a cold dense phase, but it is embedded in an envelope of more diffuse, hotter gas.

Gas metallicity is significantly above solar at several kiloparsec scales in the host galaxies even before the merger, and increases further in the galactic core of the final remnant to nearly ten times solar (Figure~\ref{fig:phase_diagrams}), consistent with observational evidence on high-redshift quasar hosts (e.g., \citealt{Wang_et_al_2022}). Note that, in previously published merger simulations, gas metallicity was assumed to be solar in the initial conditions, with little evolution taking place due to the absence of previous cosmological assembly of the target galaxy. It is therefore reassuring that a high metallicity is confirmed to be the correct starting condition when tracking the fully self-consistent cosmological evolution. The high metallicity implies efficient cooling in the optically thin regions of the galaxy, although in the core compressional heating (PdV work) is also effective \citep[see][]{Mayer_et_al_2015}, due to the supersonic gas inflows. While we do not include a transition to an optically thick radiative regime in the simulation, \citet{Mayer_et_al_2015} showed no appreciable differences between simulations that were incorporating explicitly an equilibrium temperature-density relation in the optically thick regime, and those that were not doing so, precisely because, in the core, compressional/shock heating due to both the supersonic gas inflow and gravito-turbulence was suppressing cooling in the first place.

However, the mass-weighted temperature within the central  tens of parsecs, which includes the SMD, is in the range of 10--500~K, much lower than in the original binary merger runs, in which it was of the order of 5000~K \citep[][]{Mayer_et_al_2015}. One of the reasons behind this difference is that the larger softening suppresses gravity-driven gas infall as well as shocks driven by self-gravitating spiral structures in the SMD; indeed, within the central 10~pc the softening is 5--10 times larger than the SPH smoothing length, which sets the characteristic scale of pressure \citep[see][]{Bate_Burkert_1997}. The stifling of the SMD's collapse suppresses PdV heating, which depends on the pressure gradient. Moreover, without accounting for opacity effects at high optical depth, radiative cooling is still very efficient in the SMD because of the high central density. Simply suppressing cooling  in high optical depth regions without increasing the resolution would not be the physically correct approach, though, because, as discussed in \citet{Mayer_Bonoli_2019}, the main reason behind the high core temperatures in the previous simulations was compressional/shock heating within the rapidly infalling gas. Indeed, \citet{Mayer_et_al_2015} found no difference in simulations in which a thermal balance model including various sources of opacity for high-density regions was employed and those in which optically thin cooling was used irrespective of the density range.

In addition to the fact that softening is still too large, of the order of the SMD size, to allow convergence on the state of the SMD, simple physical considerations should help us understand what should physically happen with increased resolution. Indeed, when the SMD reaches its largest mass following the gas inflow at the final merger time, $3 \times 10^8$~$M_{\odot}$, the Jeans length is as small as $\lambda_{\rm J} \sim 0.01$~pc (for a mean temperature of 100~K), which clearly implies that the SMD, had we had enough resolution, should contract to a much smaller size. Because of the high density and correspondingly high optical depth, the contraction will occur adiabatically, with the temperature evolving as $T \sim 1/R$ \citep[][]{Zwick_et_al_2023}, hence the temperature will increase to nearly $10^4$~K for a two-orders-of-magnitude decrease in disk size, which would be very comparable with the temperature observed in the binary merger simulations of \citet{Mayer_et_al_2015} with a much smaller gravitational softening.

The SMD is a differential rotator, hence Jeans collapse is just a crude indication that contraction should continue. The disk will likely become globally unstable to non-axisymmetric instabilities and then contract as a result of internal transport of angular momentum, as discussed in \citet{Zwick_et_al_2023}. With a softening comparable to the characteristic radius of a system, such instabilities are suppressed (see, e.g., \citealt{Kaufmann_et_al_2007} for a relevant numerical study in the context of bars in galactic disks). Nevertheless, we can gauge their potential importance by quantifying phenomenological stability criteria for the susceptibility to bar formation. For both uniformly rotating and differentially rotating fluids, the Kalnajs criterion $T/W$ holds \citep[e.g.,][]{BinneyTremaine1987}, where $T$ is the rotational kinetic energy and $W$ is the gravitational potential energy of the SMD. Stability to bar modes requires $T/W < 0.14$. With rotational velocities in the range 800--10000~km~s$^{-1}$ in the SMD and its vicinity, we measure $T/W \sim 0.16$--0.25 within 4~pc from the formation of the SMD to the end of the simulation. Alternatively, the susceptibility to bar instability can also be measured by the Ostriker--Peebles criterion $E_{\rm R}/T$ \citep[][]{BinneyTremaine1987}, where $E_{\rm R}$ is the random component of the kinetic energy: $E_{\rm R} = 1/2 \sigma^2$, where $\sigma^2$ is the (3D) gas velocity dispersion, of order $\sim$200--400~km~s$^{-1}$. Stability requires $E_{\rm R}/T > 5$, whereas we measured $E_{\rm R}/T < 0.5$. Therefore, we conclude strongly that, with increased resolution, a powerful bar instability would develop in the SMD.

\begin{figure}
     \centering
     \includegraphics[width = 0.49\textwidth]{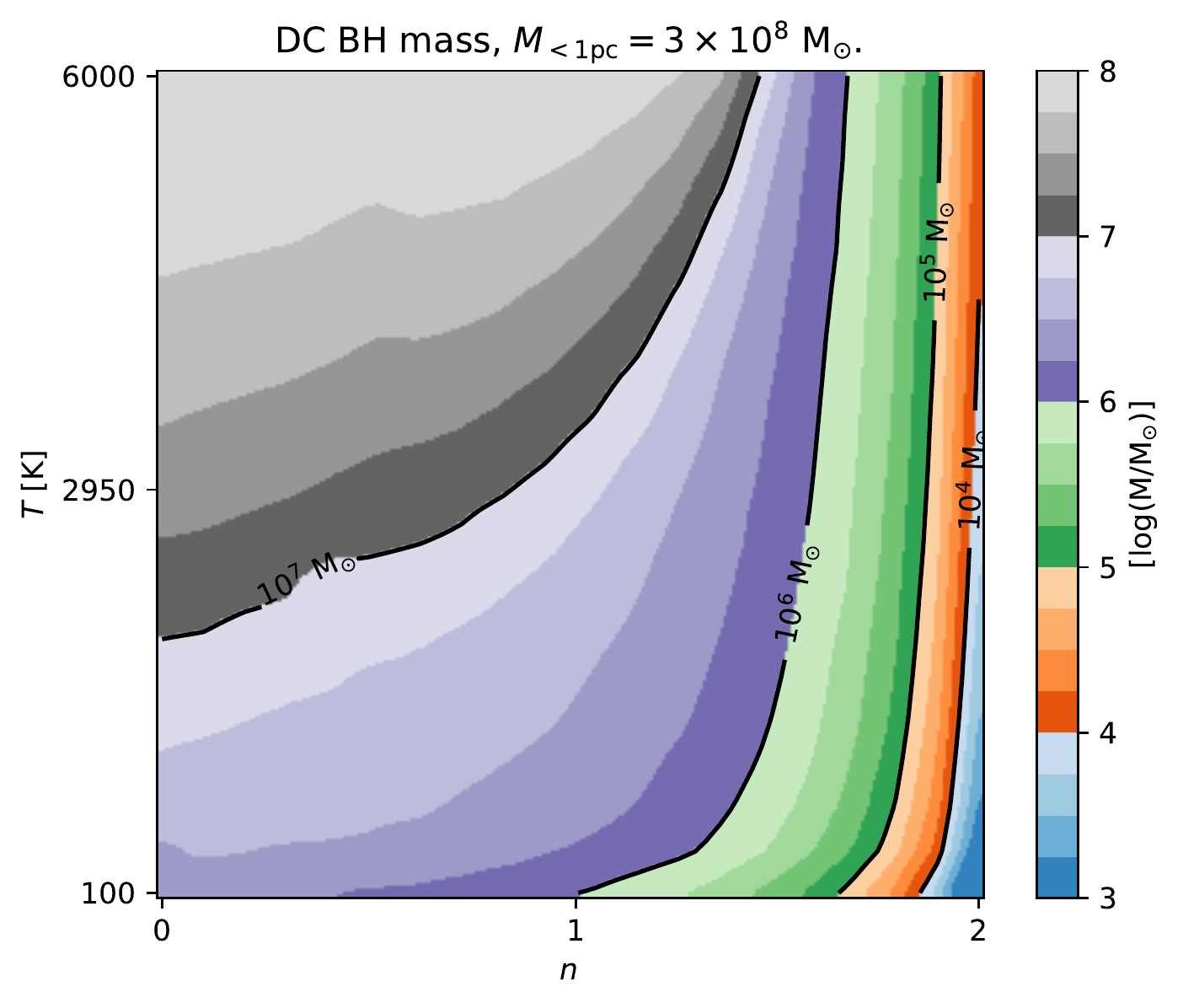}
     \caption{We show the final direct-collapse BH mass as estimated by the model of \cite{Zwick_et_al_2023}, tailored to the peak gas mass reached in the SMD of the MBPS2pc simulation after the
     galaxy merger. The results are shown as a function of the SMD initial temperature and slope of its power law density profile at sub-pc scales. As discussed in the text, exploring a range is necessary as the temperature structure of the SMD is still uncertain, while the density profile of the SMD is not resolved in the simulations below parsec scales.}
     \label{fig:final_BH}
 \end{figure}

The bar instability will induce rapid redistribution of angular momentum, increasing the central density and promoting gravitational collapse to smaller radii. In particular, contraction by 10-fold in radius is easy to achieve this way \citep[see][]{Mayer_Bonoli_2019}. We also remark that, while the SMD's dynamics is dominated by rotational support, the gas velocity dispersion is by itself also dynamically important, especially when compared to the very low temperature. Indeed, velocity dispersions of 100--200~km~s$^{-1}$ correspond to an effective kinetic temperature of $\sim$$10^6$~K, orders of magnitude higher than the actual mean gas temperature in the range 2--100~pc. Such high gas ``turbulence'' explains the resilience to fragmentation despite the extremely high densities reached in the SMD.

\subsection{Direct formation of supermassive black hole via general relativistic radial instability in a proto-supermassive star}

The physical arguments just outlined suggest that the disky core should shrink in size by about an order of magnitude. \citet{Zwick_et_al_2023} have shown that, for such a compact SMD, further dynamical or (secular) viscous transport of angular momentum will bring the system to the conditions of the global GR radial instability for a wide range of structural parameters. Therefore, we repeated the calculations in the analytical SMD evolution model of \citet{Zwick_et_al_2023}, using the maximum central gas mass reached in the isolated simulation, $3 \times 10^8 M_{\odot}$, which is about a factor of 3.5 smaller than the default range based on the numerical results of \citet{Mayer_et_al_2015}. Figure~\ref{fig:final_BH} illustrates our results on the predicted BH mass following the GR instability, as a function of the initial temperature of the SMD and of the slope $n$ of its density profile. A wide range of values for the parameter $n$ must be considered, since the slope of the SMD cannot be not constrained by the simulations, given the relatively large gravitational softening. Likewise, we let the temperature of the SMD vary from hundred to several thousand Kelvin because of the uncertain effect of stellar irradiation, which is discussed in the next section. For temperatures of a few hundred Kelvin or lower, the large majority of gas will be neutral rather than at least partially ionized, as assumed in the model by \citet{Zwick_et_al_2023}. Nevertheless, only a very small fraction of the gravitational self-energy of the SMD would be required to fully ionize the gas:

\begin{equation}
    \frac{E_{\rm{ion}} }{E_{\rm{g}}} \sim \frac{13.6 \, {\rm eV}\, \frac{M_{\rm{D}}}{m_{\rm{p}}}} {\frac{G M_{\rm{D}}^2}{R_{\rm{D}}^2}}\sim 0.3\% \times\left(\frac{ 10^8 \, M_{\odot}}{M_{\rm{D}}}\right) \left(\frac{R_{\rm{D}}}{3\, \rm{pc}} \right),
\end{equation}

\noindent where $E_{\rm{ion}}$ is the total ionization energy of an SMD with mass $M_{\rm{D}}$ and radius $R_{\rm D}$, and we neglect the effect of the disk geometry on the gravitational self-energy $E_{\rm g}$. Thus, considering that the energy budget is set by the gravitational self-energy, we expect the model to be consistent for the conditions presented in this work, provided that the SMD can start to contract adiabatically. As shown in Figure~\ref{fig:final_BH}, the mass involved in the direct collapse event strongly depends on the initial slope and temperature of the SMD. For a slope of $n\lesssim 1.5$ and a wide range of initial temperatures, the model predicts the direct formation of a BH in the range $10^6 M_{\odot}$ to $\lesssim 10^8 M_{\odot}$. This exceptionally large direct-collapse mass is a unique feature of the merger-driven scenario and, as discussed in \cite{Zwick_et_al_2023}, reproduces the expectations based on more sophisticated numerical studies of rotating SMSs under such extreme accretion conditions \citep{fowler66,baumgarteshapiro1999,haemmerle21}. For more centrally concentrated SMDs, the expected mass of the direct-collapse BH rapidly decreases as $n$ approaches 2, tending towards the more traditional heavy-seed expectation of $10^3$--$10^4 M_{\odot}$. This is an expected trend since, as discussed in \cite{Zwick_et_al_2023}, the evolution of the SMD's core more strongly resembles a typical SMS whenever when the slope parameter $n$ is close to the value 2. The formation of a fully fledged SMBH, skipping a BH seed stage, can occur whenever the slope parameter is shallow enough ($n\lesssim 1.5 $). This is a possibility that was alluded to in \citet{Mayer_Bonoli_2019} and dubbed ``dark collapse'', referring to the fact that the GR instability sets in before a typical SMS can form. In reality, this scenario would not be dark, but rather give rise to several distinctive signatures in the electromagnetic spectrum, in particular a peak flux at hard-X-ray bands. Additionally, the X-ray burst could be accompanied by an episode of neutrino emission, and by a gravitational-wave (GW) burst triggered during the asymmetric collapse \citep[which is in principle detectable by LISA; see][for a more thorough discussion]{Zwick_et_al_2023}. A forthcoming paper will explore the electromagnetic signatures more realistically, taking into account potential scattering and absorption of photons by the dense gas at larger scales that, based on the simulations, would enshroud the collapsing SMBH precursor.

\section{Discussion and Conclusions}

Our fully cosmological simulations confirm the basic scenario put forward in our previous work on merging massive primordial galaxies at $z > 8$; phenomenally intense, multi-scale gas inflows are found that accumulate enough gas in a small enough region to trigger rapid formation of either an SMS or a supermassive proto-star that could collapse directly into a massive BH by the radial GR pulsational instability, the so-called {\it dark collapse hypothesis} \citep[see][]{Haemmerle_et_al_2019,Haemmerle_et_al_2020,Haemmerle_2020}. Note that massive, centrally concentrated CNDs, which could generate secondary sub-pc scale inflows (now unresolved), are seen also in the progenitor galaxies

However, the compactness of the CNDs is significantly lower than in the SMD resulting from the merger. Indeed, assuming the enclosed mass and size of the latter, the same model by \citet{Zwick_et_al_2023} shows that the GR instability phase would not be achieved. Enhanced SF in metal-enriched gas, as seen in the simulations, is thus the correct outcome in this case.

As we have shown, the SMD is bar-unstable based on our analysis. A fast-rotating, barred SMD could lead to fission as in proto-stellar core simulations \citep[see, e.g., the review by][]{Offner2022}, producing two sub-components which might generate two comparably massive proto-SMSs, as seen in numerical relativity simulations of unstable differentially rotating polytropes \citep[e.g.,][]{Reisswig_et_al_2013}. This will produce a binary of two massive BHs and a consequent GW signal from their rapid inspiral. This GW signal would be additional to that produced by the non-axisymmetric collapse of the individual proto-SMSs into a BH, which, for large enough masses, would be detectable by LISA \citep[][]{Zwick_et_al_2023}, and, for a wide range of masses, by planned future GW experiments \citep[][]{Sesana2021}. The binary in-spiral signal would also be detectable by LISA, and, if the two BHs have masses well above $10^6 M_{\odot}$, as suggested by our results, it would be much louder than the typical MBH in-spiral signal expected at these high redshifts, because in conventional formation scenarios BH seeds would have hardly grown beyond a few million solar masses,  even with episodic super-Eddington growth phases \citep[]{Sassano_et_al_2023}. A pre-existing lighter BH seed formed by either conventional direct collapse or Pop~III stars followed by accretion would have to exceed a million solar masses to have a significant thermodynamical effect through radiative feedback \citep[]{Mayer_et_al_2015}, although the effect of AGN-driven winds and outflows is difficult to estimate. If feedback of pre-existing seeds is capable of stifling at least partially the central gas inflow, preventing the formation of a dense-enough SMD to trigger the proto-SMS formation, the conditions could still be favourable to super-Eddington accretion onto light BH seeds formed via other mechanisms \citep[see, e.g.,][]{Mayer_2019,Haemmerle_et_al_2019}.

Cautionary remarks should be made regarding the modelling of SF and radiative cooling in the very inner region of the galaxy host, where the SMD lies. SF in the higher-resolution stage of the run with particle splitting is surely overestimated. Among the most important sources of heating in the ISM subject to a starburst are the photoionization feedback from the UV output of massive stars, as well as dust heating by photoelectric effect, both absent in our simulations. This would provide significant heating, as well as modify the radiative cooling rates by affecting the abundance of ions and free electrons. Furthermore, the SMD is optically thick. Indeed, even with a mass a few times lower than in \citet{Mayer_et_al_2015}, the photon diffusion time still exceeds $10^4$~yr within 2~pc, while the dynamical time is in the range of 2000--3000~yr \citep[where we used the timescale equations as expressed in][]{Zwick_et_al_2023}. However, PdV work is less effective than in \citet{Mayer_et_al_2015}, due to both the lower inflow rates (of the order of $10^2$--$10^3$~$M_{\odot}$~yr$^{-1}$ as opposed to $> 10^4$~$M_{\odot}$~yr$^{-1}$; see Figures~\ref{fig:mollweide_10-50pc} and \ref{fig:mollweide_50-100pc}) and the  much larger softening of gravity (2~pc versus 0.1~pc). Indeed, shock heating of the nuclear region, being driven by gravitational infall, is proportional to $\sim \dot M \times {v_{\rm inf}}^2$, where $\dot M$ is the gas inflow rate and $v_{\rm inf}$ is the gas infall velocity. At scales of tens of parsecs, $\dot M$ is thus a couple of orders of magnitude smaller than in the binary mergers of \citet{Mayer_et_al_2015}, and $v_{\rm inf}$ is of the order of a few hundred km~s$^{-1}$ at a similar scale, namely a factor of a few smaller as well, hence overall, shock heating is 3--4 orders  of magnitude smaller than in \citet{Mayer_et_al_2015}. Therefore, while in \citet{Mayer_et_al_2015} the shock heating due to gravitational infall had been shown to balance out (optically thin) fine-structure line cooling and molecular cooling, thus keeping the gas isothermal at a temperature of $\sim$5000~K, here both optical depth effects reducing net radiative losses, and the heating rate by stellar irradiation (mainly photoionizing UV radiation from massive stars) would matter for the internal energy budget. The second effect is most important, though, because it operates already well  outside the SMD, where the gas is optically thin and the cooling rates in the simulation are correctly calculated because the medium is nearly optically thin.

We estimate that, with a nuclear starburst with a strength in the range 200--600~$M_{\odot}$~yr$^{-1}$ already before the galaxy merger is completed, the resulting local far-UV (FUV) heating rate per unit volume (assuming the starburst is mainly in the central kiloparsec) is $> 10^{-20}$~erg~s$^{-1}$~cm$^{-3}$. This is higher than the cooling rate provided by the most efficient coolants at $T \sim 10^3$--$10^4$~K, namely fine-structure lines (CII and OIII in particular), which provide a radiative cooling rate in the range $\sim$$10^{-25}$--$10^{-24}$~erg~s$^{-1}$~cm$^{-3}$ (these coolants are accounted for in the {\scshape CLOUDY} cooling module of the simulation). While we cannot quantify the effect of the FUV heating on the ISM temperature and  density, it is expected that the temperature of the ISM in the nucleus will be increased as well as its ionization level. As a result, a larger fraction of warm ionized gas will result in a lower SFR even before the final merger occurs, saving more mass for the gas inflow that forms the SMD. Hence, we argue that, with the photoionization heating included, the mass and temperature of the SMD would be in between what we found here and the results of \citet{Mayer_et_al_2015}. The results of this paper concerning the conditions for direct SMBH formation from the SMD should thus be viewed as conservative.

In summary, we have presented strong evidence, for the first time using cosmological simulations, that the direct formation of an SMBH via GR instability in an SMD is possible in the core of a merger between two massive disk galaxies at $z > 7$. It provides an attractive route to explain the high-redshift quasars which does not require any fine tuning of the radiative cooling properties and metallicity of the ambient medium, a clear advantage relative to conventional direct-collapse scenarios. The clustering properties, abundance, halo masses, and galaxy host metallicities are automatically in agreement with those of high-redshift quasars, because they all stem from the natural properties of high-sigma peaks in a $\Lambda$-cold-DM cosmology (see \citealt{Mayer_Bonoli_2019} for an extended discussion of this aspect). Incidentally, we note that the presence of massive  rotationally supported galactic disks at $z > 8$, which is predicted by our simulations, is in agreement with the latest observations of JWST \citep[e.g.,][]{Robertson2022}. This is an important aspect, as gravitational torques extracting angular momentum efficiently are much particularly effective in cold rotating disks. The dense gas around the site of SMBH formation gives rise to a vigorous starburst, hence the ensuing SMBH will be surrounded by a wealth of stellar-mass BHs, and possibly even intermediate-mass BHs, which would give rise to a variety of GW sources in its vicinity as they spiral-in effectively via dynamical friction and global disk-driven torques, among them extreme and intermediate mass-ratio inspirals (EMRIs and IMRIs).

\begin{acknowledgments}
The simulations were performed on the PizDaint and Alps/Eiger Cray supercomputers at the Swiss National Supercomputing Center under the uzh3 rolling allocation. The authors thank the Swiss National Science Foundation (SNF) for the support of this research under the grant 200020-192092.
\end{acknowledgments}

\bibliographystyle{aasjournal}
\bibliography{MassiveBlackPS}

\end{document}